\documentclass[11pt,preprint]{aastex}

\input epsf

\newcommand{\etal} {et al.~}  
\newcommand{\deltE}{\Delta\kern-1ptE}

\newcommand{\lam}{$\lambda$}

\def\orb[#1 #2]{{$#1^{#2}$}}
\def\tm[#1 #2 #3]{{$^{#1}{#2}_{#3}$}}

\begin{document}
 
\title{CHIANTI -- an Atomic Database for Emission Lines \\
       Paper VI:  Proton Rates and Other Improvements}
 
\author{P. R. Young,\altaffilmark{1,2}
G. Del Zanna,\altaffilmark{3}
E. Landi,\altaffilmark{4,5}
K. P. Dere,\altaffilmark{5}
H. E. Mason,\altaffilmark{3}
\and M. Landini\altaffilmark{6}}

\altaffiltext{1}{Harvard-Smithsonian Center for Astrophysics, 60
  Garden Street, Cambridge, MA 02138}

\altaffiltext{2}{Present address:
  Space Science and Technology Department, Rutherford Appleton
  Laboratory, Chilton, Didcot, Oxfordshire OX11 0QX, U.K.}

\altaffiltext{3}{Department of Applied Mathematics and
Theoretical Physics, University  
of Cambridge, Silver Street, Cambridge CB3 9EW, UK}

\altaffiltext{4}{ARTEP, Inc., Columbia, MD 21044}

\altaffiltext{5}{Naval Research Laboratory, Washington DC, 20375}

\altaffiltext{6}{Dipartimento di Astronomia e Scienza dello Spazio,
Universit\`adi Firenze, Firenze, Italy}

\begin{abstract}

The CHIANTI atomic database contains atomic energy levels, wavelengths,
radiative transition probabilities and electron excitation data for a
large number of ions of astrophysical interest. Version~4 has been
released, and proton excitation data is now included, principally for
ground configuration levels that are close in energy.
The fitting procedure for excitation data, both electrons and protons,
has been extended to allow 9 point spline fits in addition to the
previous 5 point spline fits. This allows higher quality fits to
data from close-coupling calculations where resonances can lead to
significant structure in the Maxwellian-averaged collision
strengths. The effects of photoexcitation and stimulated emission by a
blackbody radiation field in a spherical geometry on the level balance
equations of the CHIANTI ions can now be studied following modifications
to the CHIANTI software.
With the addition of \ion{H}{1}, \ion{He}{1} and \ion{N}{1}, the first
neutral species have been added to CHIANTI.
Many updates to existing ion data-sets are described, while several
new ions have been added to the database, including \ion{Ar}{4},
\ion{Fe}{6} and \ion{Ni}{21}. The two-photon continuum is now included
in the spectral synthesis routines, 
and a new code for calculating the relativistic free-free continuum
has been added. The treatment of the free-bound continuum has also
been updated.

\end{abstract}

 
\section{Introduction}

The CHIANTI database was first released in 1996 \citep{dere97} and it
contains energy levels, radiative data and electron excitation rates
for virtually all astrophysically important ions. In addition there are a
number of computer routines written in IDL which allow a user to
compute synthetic spectra and study plasma diagnostics. The database
was originally focussed towards reproducing collisionally-excited
emission line spectra at ultraviolet wavelengths from 50 to
1150\,\AA. Version 2 \citep{landi99} introduced many minor ion species
to the database as well as routines to compute free-free and
free-bound continua. The most recent version (v.3) of the database
\citep{dere01} extended coverage of CHIANTI to X-ray wavelengths
(1--50\,\AA) principally through the addition of hydrogen and
helium-like ions, and dielectronic recombination lines.

CHIANTI has seen applications to many different areas of astrophysics
since its inception. It has been extensively used in solar physics,
in particular for the analysis of spectra obtained from the CDS,
SUMER and UVCS spectrometers on board the SOHO satellite
\citep[e.g.,][]{young97, landi02, akmal01}. CHIANTI is also used to model
the instrument responses of the EIT \citep{dere00} and TRACE imaging
instruments in 
order to convert measured fluxes into physical parameters such as
temperature and emission measure. The wide  coverage of many different
ions allowed CHIANTI to be a useful aid in the verification and
definition of ultraviolet spectrometers' flux calibrations
through the 
use of emission line ratios that are insensitive to the plasma
conditions. Examples include the SERTS rocket flights \citep{young98,
brosius98}, and the Normal Incidence Spectrometer and Grazing Incidence
Spectrometers on CDS \citep{delzanna01}.

Beyond the Sun, CHIANTI has seen application to analyses of the wind
emission from the Arches cluster of massive stars \citep{raga01}, warm
gas in galaxy 
clusters \citep{dixon01} and analyses of a
number of cool stars including AB Doradus \citep{brandt01}, AU Microscopii
\citep{pagano00} and $\epsilon$ Eridani \citep{jordan01}.
\citet{delzanna02} present a review of various spectroscopic
diagnostic techniques that can be applied to XUV observations of
active stars. They use CHIANTI to illustrate the severe limitations
that some commonly-used methods and atomic data
have. \citet{delzanna02} obtain results in terms of stellar transition
region densities, emission measures and elemental abundances that are
significantly different from those of other authors. Their results
suggest that a large body of work on cool star atmospheres will have
to be revisited and stress the importance of using assessed and
up-to-date atomic data. Laboratory work also plays a vital role in the
assessment of cool star results, with work by \citet{beiers99},
\citet{brown98} and \citet{fournier01} providing valuable insights
into plasma processes affecting EUV and X-ray spectra.


CHIANTI also forms a significant part of other atomic database
packages. APED \citep{smith01} supplements CHIANTI with data from
several other sources and is focussed towards modeling X-ray
spectra. XSTAR \citep{bautista01} is a photoionization code that uses
CHIANTI data for modelling the level balance within individual ions.
CHIANTI also forms a significant part of the Arcetri Spectral Code
\citep{landi98, landl02}. 

The present work describes the latest updates to CHIANTI, including
the addition of the new physical processes of proton and photon
excitation of ion levels, the addition of new ions and revisions of
existing ion data-sets.

\section{Level balance equations}

In version 4 of CHIANTI extra processes are now included in the level
balance equations for ions, namely proton excitation and de-excitation,
photoexcitation and stimulated emission. The level balance equations
are

\begin{equation}
n_i \sum_{j \ne i} \alpha_{ij} = \sum_{j\ne i} n_j \alpha_{ji}
\end{equation}

\noindent where $i$ and $j$ are indices for the individual levels
within an ion, 
$n_i$ is the population of level $i$ relative to the population of
ions as a whole, and $\alpha_{ij}$ is the number of $i$ to $j$
transitions taking place per unit time. In previous versions of
CHIANTI, the $\alpha_{ij}$ were of the form

\begin{equation}
\alpha_{ij} = N_{\rm e} C_{ij} + A_{ij}
\end{equation}

\noindent where $A_{ij}$ is the radiative decay rate (zero if $i<j$)
and $C_{ij}$ is the electron rate coefficient such that

\begin{equation}
C_{ij} = {\omega_j \over \omega_i} \exp 
          \left( - { \deltE \over kT } \right) C_{ji} \quad i< j
\end{equation}

\noindent where $C_{ij}$ is defined in Eq.~5 of \citet{dere97},
$\omega_i$ is the statistical weight of level $i$, $\deltE$ is the positive
energy separation of levels $i$ and $j$, $k$ is the Boltzmann constant
and $T$ the electron temperature.

For version~4, $\alpha$ now takes the form

\begin{equation}
\alpha_{ij} = N_{\rm e} C_{ij} + N_{\rm p} C_{ij}^{\rm p}
+ {\cal A}_{ij} 
\end{equation}

\noindent where $N_{\rm p}$ is the proton number density, $C_{ij}^{\rm
p}$ is the proton rate
coefficient, and ${\cal A}_{ij}$ is the generalized
radiative transition rate.

\subsection{Proton rates}

\begin{figure}
\plotone{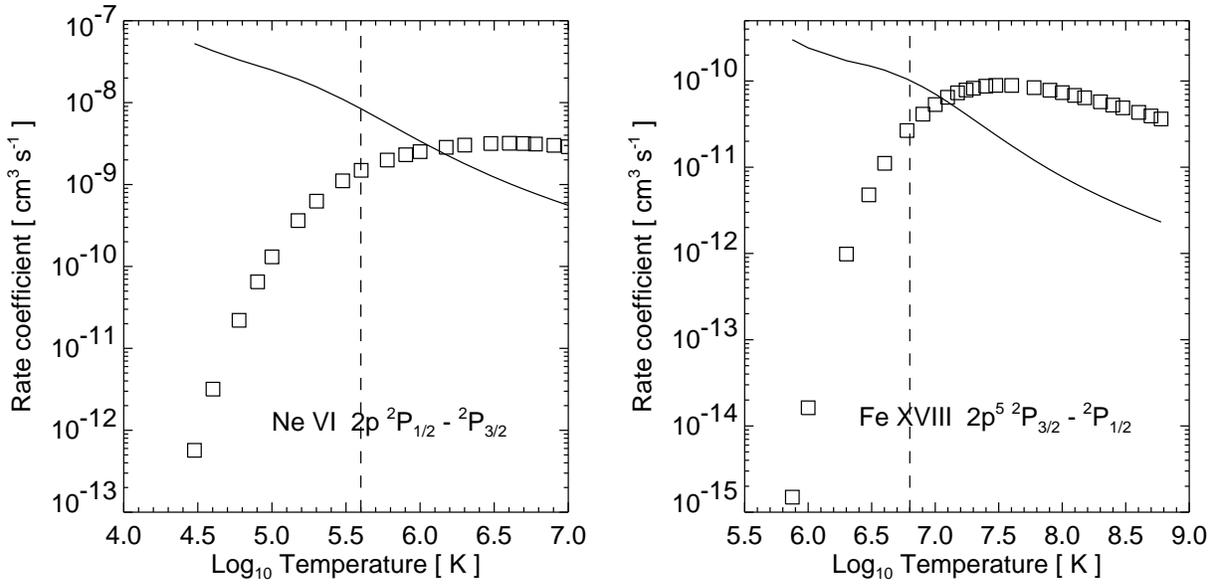}
\caption{Proton rate coefficients (squares) for the ground transitions of (a)
\ion{Ne}{6} \citep{foster97} and (b) \ion{Fe}{18} \citep{foster94b}, demonstrating the sharp fall at low
temperatures due to the collision strengths falling to  zero at
the threshold energy. For comparison, the electron rate coefficients
derived from CHIANTI are plotted as continuous curves. The dashed,
vertical lines show the $T_{\rm max}$ of the ions.}
\label{proton-plot}
\end{figure}

The inclusion of proton rates in ion level balance equations was
first demonstrated to be important in solar coronal conditions by
\citet{seaton64} for the \ion{Fe}{14} ion. He showed that the proton
rates can 
become comparable to the electron excitation rates for transitions for
which $\deltE\ll kT$. Typically only transitions within the ground
configuration of an ion are important and so, compared to the electron
processes, relatively little data are required to account for proton
processes in a particular ion.
Whereas electron collision data are usually published in the form of
collision strengths or Maxwellian-averaged collision strengths, proton
collision data are generally published directly as rate coefficients. 
As positively-charged ions repel protons, the rate coefficient falls
to zero at the threshold energy for the transition, and so tabulated
values of rate coefficients typically change by several orders of
magnitude over a small temperature range. Examples are shown in
Fig.~\ref{proton-plot} from \ion{Ne}{6} (boron-like) and \ion{Fe}{18}
(fluorine-like), where the proton rate coefficients are plotted for
the ground transitions of these ions. Also shown for comparison are
the electron rate coefficients for these transitions, derived from the data
in CHIANTI. The proton rates are seen to be comparable in strength to
the electron rates at the temperature of maximum ionization, $T_{\rm
  max}$ of the ions, and become stronger at higher temperatures. We
note, however, that the populations of the upper levels of ground
configuration 
transitions are often dominated by cascading from higher levels in the
ion rather than direct excitation \citep{mason75}.

\begin{figure}
\plotone{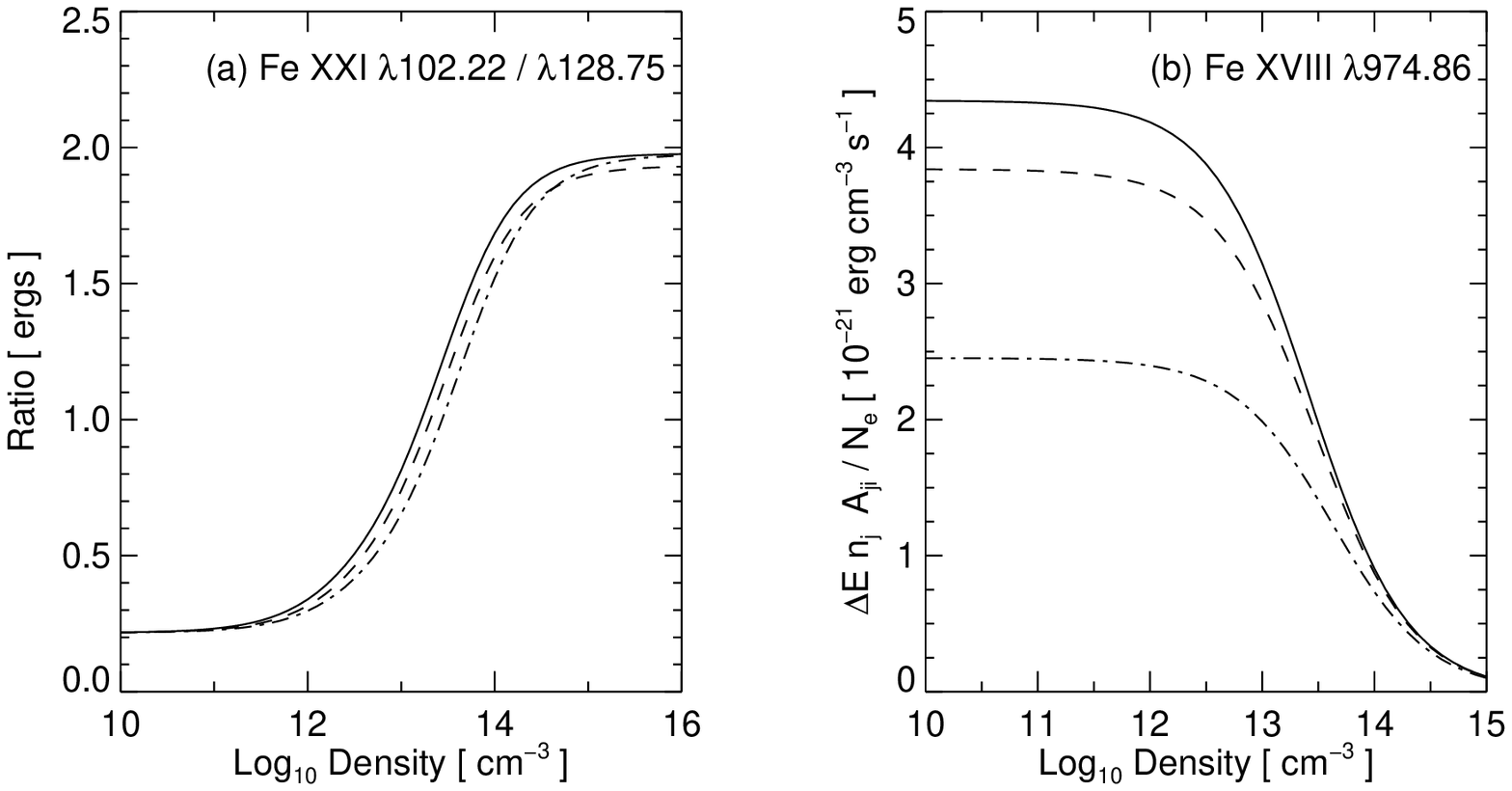}
\caption{Plot (a) compares the \ion{Fe}{21} \lam102.22/\lam128.75
  density diagnostic in the cases when proton rates are
  included (solid line) and when they are not included (dashed
  line). Also plotted is the ratio from the previous
  version of the CHIANTI database (dash-dot line). Plot (b) provides a
  similar comparison but for the \ion{Fe}{18} \lam974.86 line emissivity.
  In both cases the addition of proton rates
  and the update of the electron excitation rate result in significant
  differences.}
\label{fe18-fe21}
\end{figure}

Examples of how proton rates can affect key diagnostic emission lines
are demonstrated in Fig.~\ref{fe18-fe21}. The \ion{Fe}{18} \lam974
line is prominent in flaring, solar plasma \citep{doschek75} and has
been recently been observed in a spectrum of the star Capella
\citep{young01}. The \ion{Fe}{21} \lam102/\lam128 ratio is a key
density diagnostic for solar flares \citep{mason79} and active stars
\citep{dupree93}.

\subsection{Implementation of proton rates in CHIANTI}\label{impl-prot}

For each ion for which proton rates are available, an additional file
is required in the database to contain the fits to the rate
coefficients. The file has the suffix .PSPLUPS, and is exactly
analogous to the .SPLUPS file for the electron fits. All of the
proton transitions included in CHIANTI are forbidden transitions taking place
between levels within the same configuration, and so they are treated
as Type~2 in the \citet{burgess92} formalism. Many of the transitions
required 9-point splines (see Sect.~\ref{sect.9-point}) in order to
provide adequate
fits. The 9-point splines can only be applied when there are at least
9 points to be fitted. For some transitions with less than 9 data
points, the 5 point spline fit applied to Type~2 transitions could not
adequately fit the data. In these cases a new fit type, Type~6, was
introduced whose scaling is as follows

\begin{equation}
x={ { \left ( {kT \over \deltE} \right ) } \over
  { \left ( {kT \over \deltE} + a \right ) } }
\end{equation}

\begin{equation}
y= \log C_{ij}^{\rm p}
\end{equation}

\noindent where $a$ is the scaling parameter. By taking the logarithm
of the rate coefficient, the steep 
gradients in the rate coefficient could be overcome and reasonable
5-point spline fits obtained. As errors in the fit are amplified when
fitting logarithm data, care was taken to ensure that the rate
coefficients derived from the fits were $\lesssim$1\% from the
original data values.

For some ions the only rate coefficient data available had been
calculated at a single temperature. These data were incorporated into
CHIANTI by assuming a Type~2 transition and setting the five points of
the spline to lie on a straight 
line. The gradient of the line was set such that $y=0$ at $x=0$, and
$y=C_{ij}^{\rm p}$ at $x=0.90$. This method ensures that the rate
coefficients derived from the spline fall sharply to zero at low
temperatures, and that at high temperatures the rates tend to a
constant value that is around 10\% higher than the original data
value.

The number density of protons, $N_{\rm p}$, is required in Eq.~4 and
it is calculated from the ion
balance and element abundance files contained in CHIANTI through the
following expression

\begin{equation}
R(T) = {N_{\rm p} \over N_{\rm e}} = 
        { {\rm Ab}({\rm H}) \, F({\rm H}^+,T) \over
        \sum_{i=1}^n \sum_{Z=1}^{i} Z \,  F({\rm A}_i^{+Z}, T) \, {\rm
        Ab}({\rm A}_i)}
\end{equation}

\noindent where Ab is the element abundance, A$_i$ is the $i$th
element (i.e., A$_1$=H, A$_2$=He, etc.), $Z$ 
is the charge on the ion, and $F({\rm A}_i^{+Z}, T)$ is the fraction of
ions of element A$_i$ in the form ${\rm A}_i^{+Z}$ at temperature $T$.

The ion fractions contained in CHIANTI are tabulated over the range
$4.0\le \log\,T\le 8.0$. Above and below these values, we set $R(T)$
to the values for $\log\,T=8.0$ and $\log\,T=4.0$, respectively. The
default ion balance file used in calculating $R(T)$ is \citet{mazz98},
while the default abundances are the solar photospheric values of
\citet{grev98}.

\subsection{Photoexcitation and Stimulated Emission}

The generalized photon rate coefficient in the presence of a blackbody
radiation field of temperature $T_*$ is given by

\begin{equation}
{\cal A}_{ij} = \left\{ \begin{array}{l@{\quad}l}
                W(R) A_{ji} {\omega_j\over \omega_i} 
                {1 \over \exp(\deltE/kT_*) -1} & i<j \\
                \\
                A_{ij} \left[ 1 + W(R)
                {1 \over \exp(\deltE/kT_*) -1} \right] & i > j
                \end{array}
                \right. 
\end{equation}

\noindent where $A_{ji}$ is the radiative decay rate and $W(R)$ is the radiation
dilution factor which accounts for the weakening of the radiation
field at distances $R$ from the source center.
For a uniform (no limb brightening/darkening) spherical source with
radius $R_*$ 

\begin{equation}
W= {1 \over 2} \left[ 1 - \left( 1 - {1 \over r^2} \right)^{1/2} \right]
\end{equation}

\noindent where

\begin{equation}
r= { R\over R_*}
\end{equation}

\subsection{Implementation of Photoexcitation and Stimulated Emission
in CHIANTI}

No additions or modifications to CHIANTI data files are required for
photoexcitation and stimulated emission as their rates are entirely
determined from the radiative decay rates, level separation energies,
and statistical weights -- information already contained in CHIANTI.
It is only necessary to specify the
radiation field temperature and the dilution factor, which are done
through 
inputs to the IDL procedures with the new
keywords RPHOT and RADTEMP. RPHOT specifies $r$, the
distance from the center of the radiation source in source radii
units, while RADTEMP gives the blackbody radiation temperature in K.
By default, photoexcitation and stimulated emission are not included
in the level balance equations unless the keywords are set.

It is important to remember the assumptions in our formalism for
radiation processes. For a given ion, only very specific
wavelengths in the radiation continuum will affect the ion's level
balance. If there are significant deviations from a blackbody spectrum
at any of these wavelengths (perhaps due to a deep absorption line)
then CHIANTI may not model the ion entirely correctly.

\begin{figure}
\epsscale{0.5}
\plotone{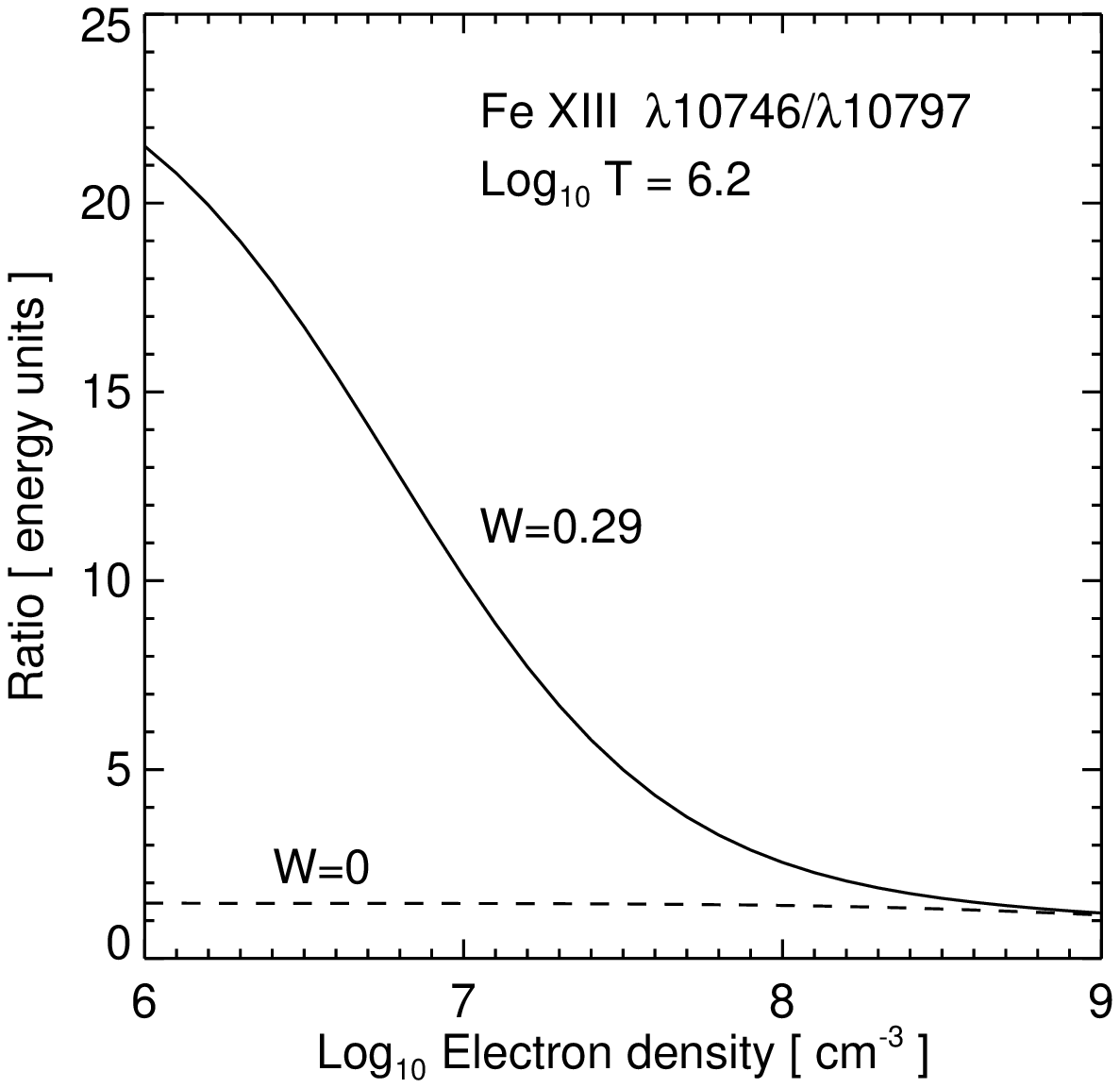}
\caption{The \ion{Fe}{13} \lam10746/\lam10797 ratio plotted as a
function of density for two different dilution factors. $W=0$
corresponds to no radiation field, while $W=0.29$ corresponds to 0.1
source radii above the source surface ($r_*=1.1$).}
\label{fe13}
\end{figure}

Examples of specific uses of the extra radiation processes include
modeling of coronal emission lines above the surface of the Sun and
other cool stars when the coronal electron density falls to low
enough values that electron collisions lose their potency. Fig.~\ref{fe13} shows the \ion{Fe}{13} \lam10746/\lam10797
ratio as a function of density, calculated in the cases of there being
no radiation field ($W=0$) and when the \ion{Fe}{13} ions are located
0.1~source radii ($W=0.29$) above the surface of a 6000~K blackbody, typical of
the Sun. The \ion{Fe}{13} infrared lines are an important density
diagnostic for ground-based solar coronal studies
\citep[e.g.,][]{penn94}, and are potential probes of the solar coronal
magnetic field \citep{judge98}.
Photoexcitation can also be important 
for modeling nebular ions that are irradiated by a hot
star, such as in planetary nebulae, symbiotic stars and Wolf-Rayet
stars. 

\section{9-point spline fitting}\label{sect.9-point}

\begin{figure}[h]
\epsscale{0.7}
\plotone{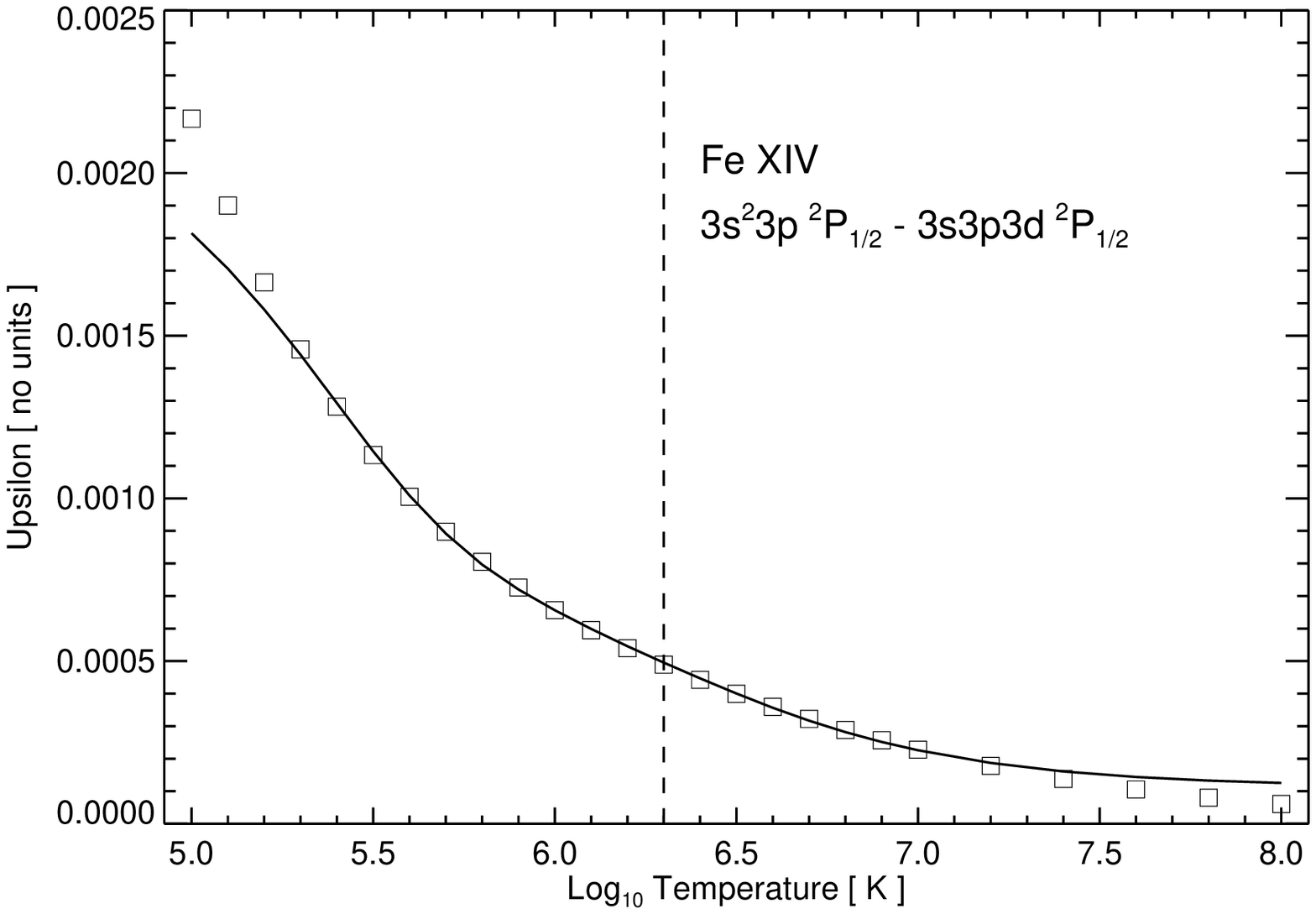}
\caption{
The squares show the $\Upsilon$ values
for the \ion{Fe}{14} $3s^2 3p$ $^2P_{1/2}$ -- $3s3p3d$
$^2P_{1/2}$ transition, calculated by \citet{smy00}. The solid line
shows the values derived from the 5-point spline fit employed in
CHIANTI, while the dashed line indicates the $T_{\rm max}$ of the
ion. At low and high temperatures the fit is seen to 
differ significantly from the original data, although good agreement is
found for temperatures close to $T_{\rm max}$.}
\label{fit-ups}
\end{figure}

Increasingly, the electron excitation data supplied by atomic
physicists are performed in  the $R$-matrix approximation which leads
to extremely  complex structure in the calculated collision strengths,
$\Omega$. These data are provided in a more convenient form as
Maxwellian-averaged collision strengths, $\Upsilon$, that are a
comparatively smooth function of temperature, and suitable for fitting
with the 5-point splines employed in the previous versions of the
CHIANTI database.  However, in some cases, it was necessary to
restrict the range of the original data in order to improve the fit
for temperatures of usefulness to astrophysical conditions. An example
is shown in Fig.~\ref{fit-ups} where one can see that at high and low
temperatures (well away from the temperature of maximum ionization,
$T_{\rm max}$, of the ion) the  $\Upsilon$ values derived from the
CHIANTI fit deviate significantly from the original data values.  An
accurate fit to both the low and high temperature data points would
require a spline with a larger number of node points.  For all of the
ions included up to and including v.3 of CHIANTI where the temperature
range of the original $\Upsilon$ data has been restricted, we have
aimed to fit within 1\%  those data points within at least 1.0 dex of
the $\log\,T_{\rm max}$ of the ion. This is  adequate for all
conditions likely to be met in stellar transition regions and coronae.

For some situations in astrophysics, particularly photoionized
plasmas, ions can be formed at electron temperatures much lower than
$T_{\rm max}$, in which case it is important to ensure accuracy of the
CHIANTI fits at lower temperatures. 
We thus now allow 9-point spline
fits to the data. In addition, the inclusion of proton rate
coefficients into CHIANTI -- described in the preceding section --
requires 9-point spline fits on account of the wide range of variation
of the rate coefficient with temperature.

The modifications made to the database and the accompanying IDL
routines in order to deal with the 9-point spline fits are described
in detail in a software note available from the CHIANTI web-page at
http://wwwsolar.nrl.navy.mil/chianti.html. No new CHIANTI files are
required; the 9-point spline fit data are incorporated into the
.SPLUPS files.

\section{The proton rate data}

Several different methods have been employed by atomic physicists in
deriving proton rate coefficients, and are reviewed in
\citet{dalgarno83} and \citet{reid88}. The most basic is the semiclassical
(or impact-parameter) approach in which the position of the proton
relative to the ion is treated classically and first order approximations are
made for the interaction with the nucleus. This approximation was
originally applied to Coulomb excitation of nuclei \citep{alder56} and
first extended to the proton excitation of ions by \citet{seaton64}.
While the first order approximation is good
for low energy collisions, at intermediate energies and low
impact parameter values the first order approximation fails, and it is
necessary to adopt a different approximation or solve numerically the
coupled differential equations describing the interaction.
Potentially the most accurate method is to treat the proton's
trajectory quantum mechanically and solve the complete set of
close-coupling equations. Such an approach is commonly used in $R$-matrix
calculations of electron collision cross-sections, however, it is
computationally much more demanding for proton collisions.
Within the semiclassical approach, it has been shown that
symmetrizing the problem with respect to the initial and final
velocities \citep{alder56}, and including polarization
effects \citep{heil82,heil83} can improve the accuracy of the proton
cross-sections.

Specific methods that have yielded the data outlined in the following
sections are as follows. 
\citet{bely70} used a symmetrized first
order semiclassical approximation to calculating the proton
cross-sections, and provided rate coefficients for a large number of
ions with configurations $2p$, $2p^5$, $3p$ and $3p^5$. In the
intermediate energy range where first order 
theory breaks down they employ an approximation referred to as
Coulomb-Bethe II, borrowed from the
theory of electron excitation of positive ions, to determine the
cross-section.

\citet{kastner79} \citep[see
also][]{kastner77} also used the first order semiclassical
approximation for low energies while for intermediate energies a form
for the cross section due to \citet{bahcall68} was used. At high
energies a further approximation to the cross-section was suggested by
\citet{kastner79}, and the combined cross-section yielded the rate
coefficients. The advantage of this method is that it can be applied
in a straightforward manner to a wide range of ions.

\citet{landman73} developed a symmetrized, semiclassical
close-coupling method to compute proton rates for \ion{Fe}{13}. This
method retains the classical treatment for the proton trajectory, but
the transition probabilities are determined by numerically solving the 
close-coupling equations, removing the uncertainties at intermediate
energies of the first order approximation.
Landman's
work was extended to a number of other ions in later papers \citep{landman75,landman78,landman80, landman79}. 

P.~Faucher used  a fully quantal close-coupling method to 
compute proton cross-sections for a number of ions \citep{faucher75,
faucher77, faucher80}. In \citet{fl77} the two authors
compared their methods for computing proton rates, and found excellent
agreement, demonstrating that the semiclassical approach is a good
approximation for highly-ionized ions at typical astrophysical
energies.

A number of papers by V.J.~Foster, R.S.I.~Ryans and co-workers have
made use of the method of  \citet{reid69} to calculate proton rate
coefficients for a large number of ions. A symmetrized, semiclassical
close-coupling 
approximation is used, with polarization effects included.

\begin{figure}[h]
\epsscale{1.0}
\plotone{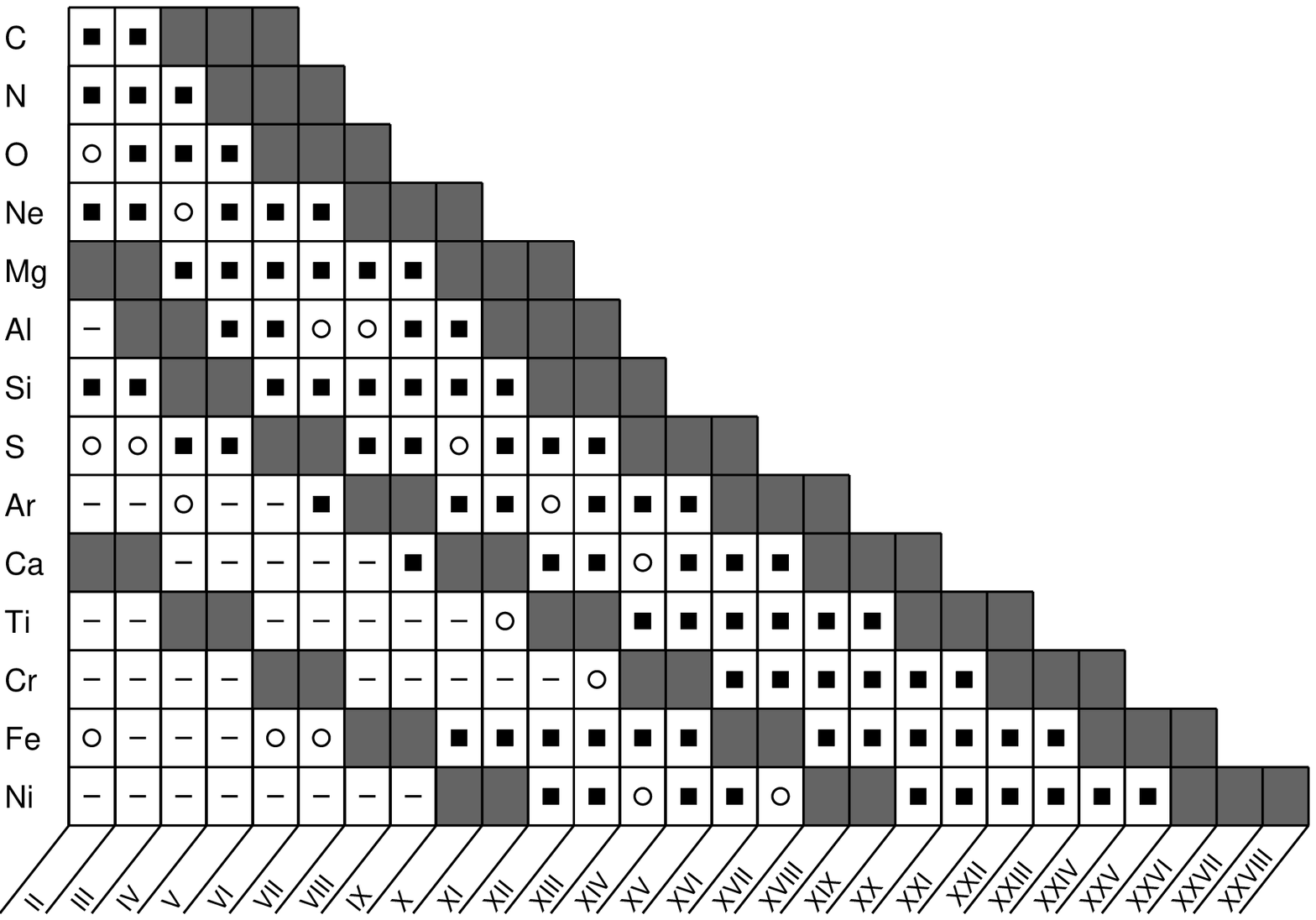}
\caption{This figure indicates for which of the major elements in
CHIANTI we have proton data. The shaded squares denote that proton
rates are unimportant for these ions; small black squares denote ions
for which data has been added to CHIANTI; circles indicate ions that
are in CHIANTI but for which no proton data is available; and dashes
indicate ions which are currently not in CHIANTI.}
\label{proton-table}
\end{figure}

Sources for most of the proton rate data assessed for CHIANTI were
obtained from the review of \citet{copeland97}, who give accuracy
ratings for each of the calculations. We have selected for each ion
those data-sets that have the highest accuracy ratings and that cover
the widest temperature range. A number of new calculations have been
published by V.J.~Foster, R.S.I.~Ryans and co-workers since this
review and have been used where available.
Ions in the hydrogen,
helium, neon, sodium, argon and potassium iso-electronic sequences all
have a single level in the ground configuration, and so proton rates
play no role in the level balance of the ions. Consequently these
sequences are not listed below. Fig.~\ref{proton-table} summarizes
which of the major elements' ions we have proton data for. Additional
data is also available for some of the minor elements (Na, P, Mn,
etc.) and these are discussed in the following sections.

\subsection{Beryllium sequence}\label{prot-be}

Rate coefficients for the $2s2p$ $^3P_J$--$^3P_{J^\prime}$ transitions have
been calculated by \citet{ryans98} using the method of \citet{reid69}
for the ions \ion{C}{3}, \ion{N}{4},
\ion{O}{5}, \ion{Ne}{7}, \ion{Mg}{9}, \ion{Al}{10}, \ion{Si}{11},
\ion{S}{13}, \ion{Ar}{15}, \ion{Ca}{17}, \ion{Ti}{19}, \ion{Cr}{21},
\ion{Fe}{23} and \ion{Ni}{25}. The rates were tabulated at 20 values
of temperature spanning typically 2--3 orders of
magnitude around the $T_{\rm max}$
of the ions. All of the data were fit with 9 point splines. In some
cases it was necessary to omit points at low and/or high temperatures
in order to obtain a good fit.

\subsection{Boron sequence}

\citet{foster97} provided rate coefficients for the ground $2s^22p$
$^2P_{1/2}$--$^2P_{3/2}$ transition and the three $2s2p^2$
$^4P_J$--$^4P_{J^\prime}$ transitions, tabulated for 20 temperatures spanning
at least two orders of magnitude. While the $^2P_{1/2}$--$^2P_{3/2}$ data
cover the temperature region $\pm 1$~dex around the
$\log\,T_{\rm max}$ of the ions, the data for the $^4P_J$--$^4P_{J^\prime}$
transitions cover a lower temperature range and generally do not
extend to 1~dex beyond the $\log\,T_{\rm max}$ of the ions. Thus care
must be taken if such data are used at temperatures well beyond the
$T_{\rm max}$ of the ion.

Data for  \ion{C}{2}, \ion{N}{3},
\ion{O}{4}, \ion{Ne}{6}, \ion{Mg}{8}, \ion{Al}{9}, \ion{Si}{10},
\ion{S}{12}, \ion{Ar}{14}, \ion{Ca}{16}, \ion{Ti}{18}, \ion{Cr}{20},
\ion{Mn}{21}, 
\ion{Fe}{22} and \ion{Ni}{24}
have been added to CHIANTI and 9 point spline fits have been
performed. For some of the transitions it was necessary to remove from
the fit
points at the beginning or end of the temperature range in order to
improve the fit quality. The cross-sections were calculated with the
method of \citet{reid69}.

\subsection{Carbon sequence}

Rate coefficients for the $2s^22p^2$ $^3P_J$--$^3P_{J^\prime}$ 
transitions have been calculated by Ryans et al.~(1999a, see also
Ryans et al.~1999b) for the ions
\ion{N}{2}, \ion{O}{3}, \ion{Ne}{5}, \ion{Mg}{7}, \ion{Si}{9},
\ion{S}{11}, \ion{Ar}{13}, \ion{Ca}{15}, \ion{Ti}{17}, \ion{Cr}{19},
\ion{Fe}{21} and \ion{Ni}{23}.
The method employed by \citet{ryans99a} is that of \citet{reid69}.
Rates are tabulated for
20 temperatures spanning 2--3 orders of
magnitude around the temperature of maximum ionization
of the ions. All of the data were fit with 9 point splines. In some
cases it was necessary to omit points at low and/or high temperatures
in order to obtain a good fit.

The inclusion of proton rates is particularly important for the heavier
ions in the carbon sequence. Fig.~\ref{fe18-fe21}a demonstrates the
effects for the \ion{Fe}{21} \lam102/\lam128 density diagnostic
ratio.

\subsection{Nitrogen sequence}

Proton rates are only available for a limited number of nitrogen-like
ions. For \ion{Ti}{16} we use the rates of \citet{bhat80c}, calculated
for nine transitions amongst the levels of the $2s^22p^3$
configuration at a
temperature of $6.5\times 10^6$~K. \citet{feldman80} give rates for
\ion{Cr}{18} and \ion{Ni}{22} for seven transitions amongst the levels of the $2s^22p^3$
configuration at temperatures of $7\times 10^6$~K (\ion{Cr}{18}) and
$1.1\times 10^7$~K (\ion{Ni}{22}). For each of \ion{Ti}{16},
\ion{Cr}{18} and \ion{Ni}{22} the method of \citet{kastner79} was used
to calculate the proton cross-sections.

Proton rates for \ion{Fe}{20} have been
calculated by \citet{bhat80a} using the method of \citet{bely70} for
all transitions amongst the levels 
of the ground configuration, tabulated at four temperatures between
$6\times 10^6$~K and $1.5\times 10^7$~K.

For \ion{Si}{8} the rates of \citet{bhat02a} have been incorporated.
These are tabulated for all transitions in the ground configuration
for nine temperatures over the range $5.6\le\log\,T\le 6.4$. 
\citet{bhat98} have calculated rates for eight transitions in the
ground configuration of \ion{Mg}{6} at a single temperature of
$\log\,T=5.6$. The four transitions from the ground $^4S$ level are
very weak and have not been included in CHIANTI. For both \ion{Mg}{6}
and \ion{Si}{8}, the method of \citet{kastner79} was used to
calculate the proton cross-sections.

\subsection{Oxygen sequence}\label{prot-o}

Unpublished data of R.S.I.~Ryans and co-workers (R.K.~Smith 1999,
private communication) for the $2s^22p^4$ $^3P_J$--$^3P_{J^\prime}$
transitions of
several oxygen-like ions have been fitted for CHIANTI. The ions are
\ion{Ne}{3}, \ion{Mg}{5}, \ion{Si}{7}, \ion{S}{9}, \ion{Ar}{11},
\ion{Ca}{13}, \ion{Ti}{15}, \ion{Cr}{17} and \ion{Fe}{19}. The proton
rates are tabulated for 20 temperatures typically spanning two orders
of magnitude around the $T_{\rm max}$ of the ion. The cross-sections
were calculated with the method of \citet{reid69}.

For the additional ions \ion{Na}{4}, \ion{Al}{6} and \ion{P}{8} the
data of \citet{landman80} are used. The proton rates are tabulated for
nine temperatures spanning a temperature interval of $\log\,T=0.8$
around the $T_{\rm max}$ of the ions, and were calculated with the
method of \citet{landman73}. Typos in the tabulation of the
proton rates for the $^3P_1$--$^3P_0$ transitions in \ion{Na}{4}
and \ion{Al}{6} at $\log\,T=5.6$ have been corrected. 

Rate coefficients for \ion{Ni}{21} have been published by
\citet{feldman80} for seven transitions amongst the levels of the
ground configuration, calculated at a temperature of $10^7$~K using
the method of \citet{kastner79}. These
single temperature data have been fit as described in
Sect.~\ref{impl-prot} and added to CHIANTI.

\subsection{Fluorine sequence}

Proton rates for the $2s^22p^5$ $^2P_{3/2}$--$^2P_{1/2}$ transition of
four F-like ions have been taken from \citet{foster94a}. The ions
are \ion{Ne}{2}, \ion{S}{8}, \ion{Ti}{14},
and \ion{Ni}{20}. The data
are tabulated for 17 temperatures spanning temperature ranges of 
$\approx$1.5~dex around the $T_{\rm max}$ of the ions, and were
calculated using the method of \citet{reid69}.

\ion{Fe}{18} was treated separately by \citet{foster94b} who tabulated
rates for 26 temperatures from $7.5\times 10^5$~K to $6\times
10^8$~K, again calculated with the method of \citet{reid69}. To obtain
a good fit to this data, the two lowest and four highest 
temperatures points had to be omitted.

Data for eight further F-like ions -- \ion{Na}{3}, \ion{Mg}{4},
\ion{Al}{5}, \ion{Si}{6}, 
\ion{P}{7}, \ion{Ar}{10}, \ion{Ca}{12} and \ion{Cr}{16} -- have been
published by \citet{bely70}. The data are tabulated for 13
temperatures spanning more than an order of magnitude around the
$T_{\rm max}$ of the ions, and have been added to CHIANTI.

An example of how the addition of proton rates affects the
\ion{Fe}{18} is presented in Fig.~\ref{fe18-fe21}b where the
emissivity of the $2s^22p^5$ $^2P_{3/2}$--$^2P_{1/2}$ ground
transition at 974.86\,\AA\ is increased by around 10\%.

\subsection{Magnesium sequence}

Proton rate coefficients for the $^3P_J$--$^3P_{J^\prime}$ transitions in the
$3s3p$ excited configuration of \ion{Si}{3},
\ion{S}{5}, \ion{Ar}{7}, \ion{Ca}{9} and \ion{Fe}{15} have been fitted
and added to CHIANTI.
The rates were calculated by \citet{landman79} and are tabulated for 11--13 temperatures at 0.1~dex intervals
around the $T_{\rm max}$ of the ions. The method for calculating the
cross-sections is that of \citet{landman73}.

\subsection{Aluminium sequence}

Proton rates are only necessary for the $3s^23p$
$^2P_{1/2}$--$^2P_{3/2}$ transition in Al-like ions.
J.~Tully~(private communication, 2001) has calculated rate
coefficients for \ion{Fe}{14} at 21 temperatures over the range
$6.0\le\log\,T\le 8.0$ and these are included in CHIANTI. The data
cover a much broader temperature range than those of \citet{heil83},
the data recommended by \citet{copeland97}. Agreement is excellent at
$1\times10^6$~K, but  the two data-sets diverge at higher
temperatures, with the Tully values being larger. This is likely due
to \citet{heil83} only calculating the cross-section to energies
$\sim$900~eV which are not large enough to obtain reliable rate
coefficients at high temperatures.

Data for a large number of Al-like ions are given in \citet{bely70},
however, most of these ions are not included in CHIANTI and so we take
data only for \ion{Si}{2}, \ion{S}{4} and \ion{Ni}{16}. The rate
coefficients are  tabulated for 13 temperatures around the $T_{\rm
max}$ of the ions.

\subsection{Silicon sequence}

Only data for \ion{Fe}{13} and \ion{Ni}{15} are available in the
literature.
For \ion{Fe}{13} we use the proton rates of \citet{landman75} that
are tabulated for five temperatures between $1\times 10^6$ and
$3\times 10^6$~K. \citet{landman75} gives data for all transitions
between the 
individual magnetic sub-levels of the $^3P_{0,1,2}$, $^1D_2$ and
$^1S_0$ levels of the ground configuration. These rates have been
summed according to Landman's Eq.~8. 

Rates for \ion{Ni}{15} are presented in \citet{faucher77} for
transitions amongst the ground $^3P$ levels. Data are given for nine
temperatures between $5\times 10^5$~K and $4.5\times
10^6$~K. The method of calculation of the proton cross-sections is
that of \citet{faucher75}.

\subsection{Phosphorus sequence}

Only data for \ion{Fe}{12} are available on the phosphorus
isoelectronic sequence.
\citet{landman78} provided rate coefficients for all transitions
within the ground $3s^23p^3$ configuration of \ion{Fe}{12},
tabulated at four 
temperatures between $1\times 10^6$~K and $2.5\times 10^6$~K and these
are included in CHIANTI.

\subsection{Sulphur sequence}

\citet{landman80} give proton rates for several members of the sulphur
sequence, but only \ion{Fe}{11} and \ion{Ni}{13} are contained in
CHIANTI. For \ion{Fe}{11} \citet{landman80} gives rates for transitions
amongst the $3s^23p^4$ $^3P_J$ ground levels, while for \ion{Ni}{13}
transitions 
from the $^3P_2$ and $^3P_1$ levels to all of the ground
configuration levels are tabulated. 

\subsection{Chlorine sequence}

For \ion{Fe}{10} and \ion{Ni}{12} we use rate coefficients for the
$3s^23p^5$ $^2P_{1/2}$--$^2P_{3/2}$ transition from
\citet{bely70} who computed data at 13 temperatures over the ranges
$5.2\le \log\,T\le 6.5$ (\ion{Fe}{10}) and $5.4\le\log\,T\le 6.7$
(\ion{Ni}{12}). \citet{bely70} also give data for a number of other
Cl-like ions, however they can not be included as no model exists for
these ions as of version~4 of CHIANTI.

Data for additional transitions in \ion{Fe}{10} were published by
\citet{bhat95} who gave rates at a single temperature of $1\times
10^6$~K for 10 transitions from the $3s^2 3p^4 3d$ $^4D$
levels. These single temperature data have been fit according to the
method described in 
Sect.~\ref{impl-prot} and added to CHIANTI. The method used to
calculate the cross-sections is that of \citet{kastner79}.

\section{New data for the standard database}

The following sections describe new electron excitation, radiative and
energy level data that have been added to CHIANTI since
version~3. Table~\ref{ioni2} summarizes which ions have been added or
updated.

\begin{table*}[!htbp]
\begin{center}

\begin{tabular}{lcccccccccccccccc}
\noalign{\smallskip}
\hline
\noalign{\smallskip}
Ion & I & II & III & IV & V & VI & VII & VIII & IX & X & XI & XII & XIII & XIV & XV & XVI \\
\noalign{\smallskip}
\hline
\noalign{\smallskip}
H  & $\star$ & \multicolumn{14}{|c}{} \\ \cline{3-3}
He & $\star$ & $\circ$ & \multicolumn{13}{|c}{} \\ \cline{4-7}
C  & & $\circ$ & $\circ$ & $\bullet$ & $\circ$ & $\circ$ & \multicolumn{9}{|c}{} \\ \cline{8-8}
N  & $\star$ & $\circ$ & $\circ$ & $\circ$ & $\circ$ & $\circ$ & $\circ$ & \multicolumn{8}{|c}{} \\ \cline{9-9}
O  & & $\bullet$ & $\circ$ & $\circ$ & $\bullet$ & $\bullet$ & $\circ$ & $\circ$ & \multicolumn{7}{|c}{} \\ \cline{10-11}
Ne & & $\circ$ & $\bullet$ & $\circ$ & $\bullet$ & $\circ$ & $\circ$ & $\circ$& $\circ$  & $\circ$  & \multicolumn{5}{|c}{} \\ \cline{12-12}
Na & & & $\circ$ & $\circ$ & $\circ$ & $\circ$ & $\circ$ & $\circ$ & $\circ$ & $\circ$ & $\circ$ & \multicolumn{4}{|c}{} \\ \cline{13-13}

Mg & & $\circ$ &          & $\circ$ & $\circ$ & $\bullet$ & $\circ$ & $\circ$ & $\circ$ & $\circ$ & $\circ$ & $\circ$ & \multicolumn{3}{|c}{} \\ \cline{14-14}


Al & &          & $\circ$ &         & $\circ$ & $\circ$ &$\bullet$& $\circ$ &$\bullet$& $\bullet$ & $\circ$ & $\circ$ & $\circ$ & \multicolumn{2}{|c}{} \\ \cline{15-15}

Si & & $\circ$  & $\circ$ & $\circ$ & $\circ$ & $\circ$ & $\circ$ & $\bullet$ & $\circ$ & $\bullet$ & $\bullet$ & $\circ$ & $\circ$ & $\circ$ & \multicolumn{2} {|c}{} \\ \cline{16-16}

P  & & & & & $\circ$ & & $\circ$ & $\circ$ & $\bullet$ & $\circ$ & $\circ$ & $\circ$ & $\circ$ & & & \multicolumn{1}{|c}{} \\ \cline{17-17}
S  & & $\bullet$ & $\circ$ & $\circ$ & $\circ$ & $\circ$ & $\circ$ & $\circ$ &$\bullet$& $\bullet$ & $\circ$ & $\bullet$ & $\bullet$ & $\circ$& $\circ$ & $\circ$ \\

Cl & & & & & & & & & & & & & & $\circ$ & &  \\

Ar & & & &$\star$ & & & $\circ$ & $\circ$ & $\circ$ & $\circ$ & $\circ$ & $\bullet$  & $\circ$ & $\bullet$ & $\bullet$ & $\circ$ \\

K  & & & & & & & & $\circ$ & $\circ$ & & $\circ$ & $\circ$ & $\bullet$ & $\circ$ & $\circ$ & $\circ$  \\

Ca & & & & & & & & & $\circ$ & $\circ$ & $\circ$ & $\circ$ & $\circ$  &   $\bullet$ & $\circ$ & $\bullet$\\

Ti & & & & & & & & & & & $\circ$ & $\circ$ & & $\circ$ & $\circ$ & $\bullet$  \\
Cr & & & & & & & & & & & & & $\circ$ & $\circ$ & & $\circ$  \\
Mn & & & & & & & & & & & & & & & $\circ$ &  \\

Fe & & $\circ$ & & & & $\star$ & $\circ$ & $\circ$ & $\circ$ & $\circ$ & $\circ$ & $\bullet$ & $\circ$ & $\circ$ & $\bullet$ & $\circ $\\
Co & & & & & & & & & & & & & & & &  \\
Ni & & & & & & & & & & & & $\circ$ & $\circ$ & & $\circ$ & $\circ$ \\
Zn & & & & & & & & & & & & & & & &  \\
\hline

\end{tabular}

\begin{tabular}{lcccccccccccc}

\noalign{\smallskip}
\hline
\noalign{\smallskip}
Ion & XVII & XVIII & XIX & XX & XXI & XXII & XXIII & XXIV & XXV & XXVI & XXVII & XXVIII \\
\noalign{\smallskip}
\hline
\noalign{\smallskip}
Ar & $\circ$ & $\circ$ &\multicolumn{10}{|c}{} \\ \cline{4-4}
K  & $\circ$ & & &\multicolumn{9}{|c}{} \\ \cline{5-5}
Ca &$\bullet$& $\circ$ & $\circ$ & $\circ$ & \multicolumn{6}{|c}{} \\ \cline{6-7}
Ti & $\circ$ & $\circ$ & $\circ$ & $\circ$ & & & \multicolumn{5}{|c}{} \\ \cline{8-9}
Cr & $\circ$ & $\bullet$ & $\circ$ & $\circ$ & $\circ$ & $\circ$ & & & \multicolumn{4}{|c}{} \\ \cline{10-10}
Mn & $\circ$ & $\circ$ & $\bullet$ & $\circ$ & $\circ$ & $\circ$ & $\circ$ &         &          & \multicolumn{3}{|c}{} \\ \cline{11-11}
Fe & $\circ$ & $\bullet$ & $\bullet$ &$\bullet$& $\bullet$ & $\bullet$ & $\bullet$ & $\circ$ & $\circ$  & $\circ$  & \multicolumn{2}{|c}{} \\
\cline{12-12}

Co & $\circ$ &         & $\circ$ &         & $\bullet$ & $\circ$ & $\circ$ & $\circ$ & $\circ$ & & & \multicolumn{1}{|c}{} \\ \cline{13-13}

Ni & $\circ$ & $\circ$ & $\circ$ & $\circ$ &  $\star$  &  $\bullet$ &  $\circ$ & $\circ$ & $\circ$ & $\circ$ & $\circ$  & $\circ$ \\
Zn &         &         &         & $\circ$ & & & & $\bullet$ & $\circ$ & $\circ$ & $\circ$ \\
\noalign{\smallskip}
\hline
\normalsize
\end{tabular}
\caption{Ions included in the CHIANTI database.
$\circ$: Ions in the CHIANTI 3.0 version {\em not} changed in the
present update. 
$\bullet$: Ions in the CHIANTI 3.0 version whose data have been
modified/complemented in the 
present update. $\star$: New entries for the CHIANTI  version 4.0
database.}
\label{ioni2}
\end{center}
\end{table*}

\subsection{Hydrogen isoelectronic sequence}

\subsubsection{H\,I}

The atomic model for \ion{H}{1} includes the 25 fine structure levels
of the
$1s$, $2l$, $3l$, $4l$ and $5l$
configurations.
Observed energies are taken from the National Institute of Science and
Technology (NIST) Atomic Spectra Database \citep{nist}.  For oscillator
strengths and radiative decay rates ($A$ values) of allowed
lines, the values of \citet{wiese66} have been used.  The magnetic
dipole and two photon decay rates from the first excited level $2s$
$^2S_{1/2}$ are taken from \citet{parpia82}.

Maxwellian-averaged collision strengths are taken from the $R$-matrix
calculations of \citet{anderson2} who consider the 15 LS levels up to
$5l$.  The orginal calculations of \citet{anderson1} had been
assessed for inclusion in CHIANTI but it was found that their scaled
collision strengths for allowed transitions did not approach the high
temperature limit specified by \citet{burgess92}.  This was brought to
the attention of the authors who found an error in their treatment of
the $R$-matrix calculations at high energies.  \citet{anderson2} report
the revised collision strengths.  Fine structure collision strengths
are derived under the assumption of LS coupling.  \citet{aggarwal91}
previously calculated \ion{H}{1} collision strengths, also using $R$-matrix
methods, but at lower energies and temperature than the Anderson
calculations.  In general, the two sets of calculations are in
reasonable agreement at temperatures near 10$^4$~K but then tend to
diverge at higher temperatures where differences on the order of a
factor of 2 can often be found.

\subsection{Helium isoelectronic sequence}

\subsubsection{He\,I}

For the helium isoelectronic sequence, the 49 fine structure levels of
the $1snl$ configurations, $n$=1--5 and $l=s,p,d,f,g$
are
included.  Observed energies are taken from the NIST Atomic Spectra
Database \citep{nist}. 
For oscillator strengths and radiative decay rates ($A$
values) of allowed lines, the values of \citet{wiese66} have been
used.  The magnetic dipole transition probabilities for $1s$ $^1S_0$ --
$2s$ $^3S_1$, $1s$ $^1S_0$ -- $2p$ $^3P_{1,2}$ and $2s$ $^3S_1$ -- $2p$
$^3P_{0,1,2}$  are taken from the calculations of \citet{lin}.  The
two photon decay rate for $1s$ $^1S_0$ -- $2s$ $^1S_0$ is taken from
\citet{drake86}. 

\citet{sawey} have calculated collision strengths for \ion{He}{1} for all of
the levels included in the current CHIANTI model.  Maxwellian-averaged
collision strengths between temperatures of 2000 and 30,000~K are
provided.  Since the population of He I peaks at a temperature of about
30,000~K under conditions of coronal ionization equilibrium
\citep{mazz98}, this range of temperatures is not sufficient for most
diagnostic applications.  Recently, \citet{bray} have calculated He I
collision strengths at temperatures between 5600 and 560,000~K using
the convergent close-coupling method.  When the two calculations are
compared, there is often, but not always, a reasonable agreement in
the temperature region where the two calculations overlap.  For the $1s^2$ --
$1snp$ allowed transitions, the Bray calculations tend to the high
temperature limit in a smooth manner.  However, this does not appear to
be the case for all allowed transitions.  In combining the two sets of
calculations, we have included all of the \citet{sawey} collision
strengths and the highest temperature value of the \citet{bray}
values.  In our usual manner, the scaling laws of \citet{burgess92} have
been applied to the combined set of collision strengths and splines fit to the scaled collision strengths.
Previously, \citet{lanzafame} showed how the \ion{He}{1} calculations could be
extended to higher temperature by applying the same scaling laws for
the allowed transitions.

\subsection{Lithium isoelectronic sequence}

\subsubsection{C\,IV, O\,VI}

The collisional data for the $n=2,3,4$ configurations have been
replaced with the $R$-matrix calculations from \citet{griff00}.  These
correspond to the 15 lowest energy levels in the atomic model for both
ions.

The calculations were carried out using the $R$-matrix with
pseudo-states method \citep[RMPS][]{bart96}, including 9 spectroscopic
terms of the configurations \orb[1s 2]\orb[2s ], \orb[1s 2]\orb[3s ],
\orb[1s 2]\orb[3p ], \orb[1s 2]\orb[3d ], \orb[1s 2]\orb[4s ], \orb[1s
2]\orb[4p ], \orb[1s 2]\orb[4d ]  and \orb[1s 2]\orb[4f ], and 32
pseudo states \orb[1s 2]\orb[nl ] for $n=5$ to 12 and $l=0$ to
3. Results showed that the presence of pseudo states affects mostly
transitions to the $n=4$ levels, and is less important for the transitions
to lower levels.

Effective collision strengths are calculated in LS coupling, and they
have been  scaled into intermediate coupling by using the statistical
weights of the levels.  Effective collision strengths are provided by
\citet{griff00} in the $4.2\le \log T\le6.5$ temperature range for
\ion{C}{4}, and in the $4.55\le\log T\le6.85$ temperature range for
\ion{O}{6} ($T$ in K).

The radiative and collisional data for additional transitions remain
unchanged.

\subsection{Beryllium isoelectronic sequence}

In the earlier versions of the database, the collisional data for the
$n=2$ levels in the Be-like sequence were generally taken from  the
distorted wave calculations from \citet{zhang92}. The only exceptions
were \ion{Ne}{7},  \ion{Mg}{9} and the minor ions, for which Version~3
of the database adopted $R$-matrix results. However, it has been
found that resonances play an important role in the calculation  of
Be-like $n=2$ effective collision strengths \citep{landi01}, so the
distorted wave data  for the $n=2$ transitions in the whole sequence
have been replaced with close coupling results, as described
below. The accuracy of the $R$-matrix data in the case of \ion{Ne}{7}
was demonstrated from laboratory spectra by \citet{mattioli99}.

\subsubsection{O\,V, Si\,XI}

C. Jordan (2001, private communication) noted that the \ion{O}{5}
\lam1218/\lam1371 line ratio calculated  
with the Version~2 distorted wave rates provided unusually high
densities for the Sun, in 
disagreement with values from other ions formed at similar
temperatures. The use of the  
\citet{berr85} $R$-matrix results yielded more realistic density values,
and so these 
data have been adopted in the present version of the database both for \ion{O}{5} and \ion{Si}{11}. 

The data consist of LS coupling effective collision
strengths calculated for all transitions between the six $n=2$ terms
using the $R$-matrix 
method. Intermediate coupling effective collision strengths were
obtained by scaling 
the Berrington \etal (1985) data with the statistical weights of the
levels. Data are provided  
for the temperature range $4.5\le\log T\le6.1$ for \ion{O}{5} and
$5.4\le\log T\le7.0$
for \ion{Si}{11}. Although it is not easy to assess the quality of
such calculations,  
Berrington \etal (1985) claim an accuracy of 10--20\% for their results.

All other data for these two ions remain unchanged.

\subsubsection{Al\,X}

Collisional data for \ion{Al}{10} have been taken from
\citet{keenan86}, who interpolated the $R$-matrix data of
\citet{berr85} for \ion{C}{3}, 
\ion{O}{5}, \ion{Ne}{7} and 
\ion{Si}{11}.  Effective collision strengths are provided for transitions
between all of the 10 fine structure $n=2$ levels in the \ion{Al}{10}
model. \citet{keenan86} claim
that their interpolated data are accurate to within 10~\% over a
temperature range of 
$\pm$0.08 dex from the maximum \ion{Al}{10} fractional abundance in
ionization equilibrium, 
corresponding to a temperature range of $5.3\le\log T\le6.9$ ($T$ in K).

All other data for \ion{Al}{10} remain unchanged.

\subsubsection{S\,XIII, Ar\,XV}

Collisional data for \ion{S}{13} and \ion{Ar}{15} have been taken from
\citet{keenan88} who interpolated $R$-matrix data for \ion{Ne}{7}, \ion{Si}{11}
\citep{berr85} and 
\ion{Ca}{17} \citep{dufton83}. Effective collision strengths are
provided for transitions 
between all of the 10 fine structure $n=2$ levels. \citet{keenan88} claims that
these interpolated data are 
accurate to within 10\% over a temperature range of $\pm$~0.08 dex from
the maximum fractional  
abundance of each ion, corresponding to $5.6\le\log T\le7.2$ for
\ion{S}{13} and $5.7\le\log T\le7.3$ for \ion{Ar}{15}.
\citet{keenan88} notes that for some transitions his results are
significantly different 
from the distorted wave results of \citet{bfs86}.

All other data for these ions remain unchanged.

\subsubsection{Ca\,XVII}

\ion{Ca}{17} collisional data for the $n=2$ levels have been changed,
in order to use the $R$-matrix results from
\citet{dufton83}. Effective collision strengths were calculated in LS
coupling for all of the $n=2$ levels, and then converted into
intermediate coupling collisional  data. At low impact energies, an
extension of the $R$-matrix method was applied to take into account
relativistic effects in the scattering equations.  Effective collision
strengths were provided in the $6.4\le\log T\le 7.2$ temperature
range.  Oscillator strengths are also provided by \citet{dufton83},
and these have been used to scale the effective collision strengths
according to the \citet{burgess92} scaling  laws.
\citet{dufton83} finds good agreement between his $R$-matrix results
and the distorted  wave calculations of \citet{bhat83}.

All other data for \ion{Ca}{17} remain unchanged.

\subsubsection{Fe\,XXIII}

\ion{Fe}{23} distorted wave collision rates for the $n=2$ levels have
been replaced by the $R$-matrix calculations carried out by
\citet{chid99} as part of the Iron Project. Effective collision
strengths for the transitions were tabulated over the
temperature range $6.3\le\log T\le 8.1$.

\citet{chid99}  have compared their results with the distorted wave
calculations from \citet{zhang92} and \citet{bhat86}, finding that the
background collision strengths  were in good agreement in most
cases. However, the neglect of resonances in the distorted wave data
leads to large differences in some of the effective collision strengths.

\subsection{Boron isoelectronic sequence}

\subsubsection{Al\,IX, Si\,X, S\,XII, Ar\,XIV, Ca\,XVI}

Earlier versions of the CHIANTI database adopted the $R$-matrix
calculations of \citet{zhang94} for transitions within the $n=2$
levels.  However, a small error has been found in the data (H.L.~Zhang
2001, private communication), whose effect is   small, but
non-negligible for ions heavier than \ion{Al}{9}.  \citet{keenan00}
published revised collisional rates for \ion{Si}{10} and compared them
with the observations of a solar active region, demonstrating the
accuracy of the  new data.

In Version~4 of the database, we have adopted the revised values of
the $n=2$ electron excitation rates kindly made available to us by
Dr. H.L.~Zhang, for all of the most abundant ions in the sequence that
are affected by the correction in the calculation: \ion{Al}{9},
\ion{Si}{10}, \ion{S}{12}, \ion{Ar}{14} and \ion{Ca}{16}.  The data
were calculated using the $R$-matrix method for a large range of
temperatures, from $T/z^2$=100 to 50000  ($T$ in K), where $z=$8
(\ion{Al}{9}), 9 (\ion{Si}{10}), 11 (\ion{S}{12}), 13 (\ion{Ar}{14})
and 15 (\ion{Ca}{16}).

All other data are left unchanged.

\subsubsection{Fe\,XXII}

The $R$-matrix calculations from \citet{zhang94} have been superseded by
an extensive computation carried out by \citet{bad02}. The atomic
model includes a total of 20 configurations, giving rise to 204 fine
structure levels. The configurations included are \orb[2s 2]\orb[2p ],
\orb[2s2p 2], \orb[2p 3],  \orb[2s 2]\orb[nl ], \orb[2s2pnl ] and
\orb[2p 2]\orb[nl ] where $n=3$ and $l=s,p,d$; and $n=4$ and
$l=s,p,d,f$.

Observed energy levels are taken from \citet{shirai00} and
\citet{kelly87}. Theoretical energy levels,  radiative decay rates and
effective collision strengths for all the levels in the atomic model
come from \citet{bad02}. The observed energies of a few levels have
been interchanged to match the ordering of the theoretical
energies. Einstein coefficients for a few forbidden transitions within
the ground  configuration, not available in \citet{bad02}, were taken
from \citet{gal98}; a comparison  between the \citet{bad02} and
\citet{gal98} radiative rates for common transitions shows that  the
two calculations agree to within 10\%.

\citet{bad02} carried out $R$-matrix calculations of collision strengths
using the Intermediate Coupling Frame Transformation method
\citep[ICFT;][]{griff98} for all possible transitions in the atomic
model.  Effective collision strengths are provided for temperatures in
the range $4.99\le\log\,T\le 6.99$.  However,
\citet{bad02} warn that data for $n=3$ and 4 levels at temperatures
lower than  2.4$\times 10^6$~K are less reliable, and so these low
temperature data have not
been considered for inclusion in the database. The \citet{bad02} dataset is the most extensive
available in the literature; we also note that \citet{zhang97a} have
carried out a relativistic $R$-matrix calculation for collisional
excitation rates for the first 45 levels in the \ion{Fe}{22} model as
part of the Iron Project.

\subsection{Carbon isoelectronic sequence}

\subsubsection{Ne\,V}

Radiative and collisional data for 49 fine structure levels of
\ion{Ne}{5} have been calculated by \citet{griff00a} using the ICFT
method within the $R$-matrix approximation. 46 levels belong to the
$2s^22p^2$, $2s2p^3$, $2p^4$, $2s^22p2l$ ($l=s,p,d$), and the
remaining levels belong to 
the $2s2p^23s$ $^5P$ term. The effective collision strengths cover
the temperature range $3.0\le\log\,T\le 6.0$.

\citet{griff00a} did not provide the forbidden radiative decay rates
for the ground configuration and so we have adopted the data provided
by \citet{bhat93}, except for the $^3P_{1,2}$--$^1D_2$ transitions
which come from \citet{storey00}.

Experimental energy levels are available for all 49 fine structure
levels in the model; $n=2$  
level energies are taken from \citet{edlen85}, while for all other
levels the values from the NIST 
database \citep{nist} are used.

\subsubsection{Fe\,XXI}

The \ion{Fe}{21} atomic model in Version 3.02, that included 68 fine
structure levels arising from 9 different configurations, has been
increased to 290 fine structure levels from 18 configurations. These
are the \orb[2s 2]\orb[2p 2], \orb[2s2p 3], \orb[2p 4], \orb[2s
2]\orb[2p3l ], \orb[2s2p 2]\orb[3l ], \orb[2p 3]\orb[3l ] ($l=s,p,d$),
\orb[2s 2]\orb[2p4l ] ($l=s,p,d,f$) and \orb[2s 2]\orb[2p5l ]
($l=s,d$) configurations. Experimental energies are available only for
a few levels and their values come from \citet{shirai00}; additional
energies are taken from \citet{mason79}, \citet{bromage77},
\citet{kelly87} and the laboratory
measurements of \citet{brown02}.

The \ion{Fe}{21} dataset comes from three different sources, each
providing a complete set of theoretical energy levels, radiative
coefficients and collision rates. Data for the levels belonging to the
\orb[2p 3]\orb[3l ]  configurations are taken from \citet{zhang97b},
data for $n=5$ configurations come from \citet{phill96}, while for all
the other levels we have adopted the recent \citet{bad01} dataset.

\citet{bad01} provided radiative data for all possible transitions
within the lowest 200 levels of the present \ion{Fe}{21} model, with
the only exceptions the forbidden transitions between levels in the
ground configuration; for these, radiative data have been calculated
using the program SSTRUCT \citep{eiss74}. The \citet{bad01} collision
rates have been calculated using the $R$-matrix method in conjunction
with the ICFT method; effective collision strengths are provided in
the temperature range $4.9\le\log\,T\le 7.9$. However, due to the
uncertainties in the calculation of the $n>2$ level energies,
effective collision strengths are recommended only for temperatures
greater than $2\times 10^6$~K.

Both \citet{zhang97b} and \citet{phill96} calculate collision
strengths under the distorted wave approximation; both authors also
provide data for many more levels and transitions, for which results
from \citet{bad01} are also available. As noted by \citet{bad01},
resonances play a major role in the collisional excitation rates for
these levels, and consequently $R$-matrix results are considered more
accurate and have been preferred.

Recently, also \citet{butler00} carried out extensive $R$-matrix
calculations for \ion{Fe}{21} as part of the Iron Project, including
$n=2$ and $n=3$ levels. However,  the larger target representation
adopted by \citet{bad01} allows the inclusion of important  resonances
that significantly affect several $n=2 \to n=2,3$ transitions, and so
the latter results have been preferred. On the contrary, comparison
with results from \citet{phill96} for the $n=4$ levels  shows good
agreement, demonstrating that resonances are unimportant for
transitions from these levels.

Fig.~\ref{fe18-fe21}b demonstrates the difference between the
\ion{Fe}{21} \lam102/\lam128 density diagnostic ratio calculated with
this new model of the ion (with and without proton rates), and with
the previous version of CHIANTI.

\subsection{Nitrogen isoelectronic sequence}

\subsubsection{N\,I}

The CHIANTI atomic model for \ion{N}{1} includes 26 fine structure levels arising from 4 different 
configurations: \orb[2s 2]\orb[2p 3], \orb[2s2p 4], \orb[2s 2]\orb[2p 2]\orb[3s ] and 
\orb[2s 2]\orb[2p 2]\orb[3p ]. Experimental energies come from the NIST database \citep{nist}
and are available for all 26 levels.

Theoretical energy levels and radiative data come from the
calculations of \citet{hibb91}, carried out 
using the CIV3 code of \citet{hibb75}. \citet{hibb91} provide $A$ values for transitions between
the ground and excited configurations, and between two excited configurations, in intermediate 
coupling. Data for forbidden transitions within the ground configurations were taken from
the NIST database. It is to be noted that \citet{tayal99} calculated oscillator strengths
for all transitions among the \orb[2s 2]\orb[2p 3], \orb[2s2p 4], \orb[2s 2]\orb[2p 2]\orb[3l ] 
($l=s,p,d$) configurations, but provided only LS coupling results. However, \citet{tayal99}
report good agreement between their results and earlier calculations, including \citet{hibb91}.

\citet{tayal00} calculated fine-structure effective collision strengths for transitions within the
ground configuration and from the ground to the excited configuration levels in the CHIANTI model.
\citet{tayal00} adopted the $R$-matrix approximation, including 18 LS states in the target representation.
Data are provided in the temperature range $3.0\le \log\,T\le 5.75$.

It is to be noted that the effective collision strengths of the \orb[2s 2]\orb[2p 3] \tm[4 S 3/2] --
\orb[2s 2]\orb[2p 2]\orb[3p ] \tm[2 D 3/2] and \tm[2 D 5/2] transitions tabulated in \citet{tayal00} 
were incorrect; the author has kindly provided us revised values.

\subsubsection{O\,II}

All radiative data for the ground $2s^22p^3$ configuration of
\ion{O}{2} have been updated with the calculations of
\citet{zeipp87}. We note that two errors in the earlier model of
\ion{O}{2} have been corrected: the $A$ value for the $2s^22p^3$
$^4S_{3/2}$--$^2D_{3/2}$ transition had inadvertently been assigned to
the $2s^22p^3$
$^4S_{3/2}$--$^2D_{5/2}$ transition, and vice versa. The same error
also occurred for the $2s^22p^3$
$^4S_{3/2}$--$^2P_{1/2}$ and $^4S_{3/2}$--$^2P_{3/2}$ transitions. The
$^4S$--$^2D$ transitions give rise to the prominent density diagnostic
ratio \lam3726/\lam3729 for nebular plasmas \citep{seaton57}, and the
corrections to 
the CHIANTI model now give excellent agreement to earlier work on this
ratio \citep[e.g.,][]{stang89}.

\subsubsection{Mg\,VI, Al\,VII, P\,IX, K\,XIII, Ca\,XIV, Cr\,XVIII,
  Mn\,XIX, Co\,XXI, Ni\,XXII}\label{n-zhang}

The CHIANTI models for these ions have been updated with the data of
\citet{zhang99} who provide collision strengths, theoretical energy levels
and oscillator strengths for all transitions between the 15 levels of the
$2s^22p^3$,
$2s2p^4$ and $2p^5$ configurations, increasing the number of
transitions predicted by the CHIANTI models. The collision strengths were
calculated at six values of the incoming electron energy, and 5 point
spline fits were
performed to the 105 transitions of each ion to an accuracy of
$\lesssim 1$\%.

$A$ values for the forbidden transitions in the ions have been taken
from \citet{merkelis99} where available, and \citet{zhang99}
otherwise. Oscillator strengths and $A$ values for allowed transitions
are from \citet{zhang99}. Experimental values of the 15 energy levels
of each ion were obtained from \citet{edlen84}.

We have compared the radiative data 
with those present in the previous version of CHIANTI and the few
values 
available in the  NIST database \citep{nist}, and have found only small 
differences (within 10--20\%) for most transitions. In earlier
versions of CHIANTI the data for \ion{Al}{7}, \ion{P}{9}, \ion{K}{13}, \ion{Cr}{18},
\ion{Co}{21} and \ion{Ni}{22} 
had been derived through interpolation of the
data-sets of neighbouring ions on the isoelectronic sequence \citep{landi99}.
The good agreement of the data-sets confirms the validity of the interpolation 
procedure adopted in v.2 of CHIANTI.

For \ion{Mg}{6}, the additional data for the $2s^22p^23s$
configuration of \citet{bhat98} described in \citet{landi99} have
been retained.

\subsubsection{Si\,VIII}

The new CHIANTI model for \ion{Si}{8} contains all 72 levels of the
$2s^22p^3$, $2s2p^4$, $2p^5$ and $2s^22p^23l$ ($l=s,p,d$)
configurations.
For transitions amongst the levels of the ground $2s^22p^3$
configuration, the $R$-matrix data of \citet{bell01} are used. The
authors tabulated Maxwellian-averaged collision strengths for temperatures
$3.3\le\log\,T\le 6.5$, and these data were fitted with nine point
splines. In order to achieve an accuracy of $\lesssim 1$\% in the fits
it was necessary to omit the two lowest temperature data points, and
so the fits apply to the temperature range $3.7\le\log\,T\le 6.5$.
The new $R$-matrix data modify the \ion{Si}{8} ground configuration
line emissivities by 
$\sim 10$\% compared with the \citet{zhang99} distorted wave data.
For all remaining $n=2$ transitions, the distorted wave collision
strengths of \citet{zhang99} were used.

\citet{bhat02a} have calculated collision strengths for transitions
involving the $n=3$ levels using the University College of London
distorted wave programs \citep{eiss72,eiss98}. The collision strengths
were calculated at  incident electron energies of 20, 40, 60 and
80~Ry.

Radiative data for forbidden transitions are from \citet{merkelis99}
and all other $n=2$ radiative data are from \citet{zhang99}. For
transitions involving the $n=3$ levels, radiative data are from
\citet{bhat02a}.
Experimental energies for the $n=2$ levels are from \citet{edlen84},
while $n=3$ energies have been compiled from the NIST database \citep{nist} and
\citet{kink99}. For a number of the $n=3$ levels experimental energies
were unavailable and for these the theoretical values of
\citet{bhat02a} have been used.

The extension of the CHIANTI \ion{Si}{8} model to include the $n=3$ levels
is particularly important as a number of $n=3$ to $n=3$ transitions
have been identified in solar and laboratory spectra between 900 and
1300\,\AA\ \citep{kink99}. A number of ground forbidden transitions
are also found in this wavelength range, and ratios between the two
sets of lines have considerable diagnostic potential.

\subsubsection{S\,X}

\citet{bell00} have published Maxwellian-averaged collision strengths
for all 231 possible transitions amongst the 22 levels of the
$2s^22p^3$, $2s2p^4$, $2p^5$ and $2s^22p^23s$ configurations,
calculated in the $R$-matrix approximation. The values are tabulated
for 12 temperatures over the range $4.6\le\log\,T\le 6.7$ and have been
fitted with 5 point or 9 point splines to yield fits accurate to
$\lesssim 1$\%. For five transitions, errors were found in the
original tabulated data and have been corrected. More details are
provided in the comments section of the CHIANTI .SPLUPS file.

Radiative data are from \citet{merkelis99} for the forbidden
transitions and \citet{zhang99} for all other $n=2$ transitions. No
data were available in the literature for the $n=3$ levels and so
SSTRUCT was used to derive oscillator strengths and decay
rates. The model used contained 24 configurations and showed excellent
agreement with the data of \citet{merkelis99} and \citet{zhang99} for
the $n=2$ levels. 

Experimental energies are available for all 22 levels and we use the
values of \citet{edlen84} for the $n=2$ levels, and the values from
the NIST database \citep{nist} for the $2s^2 2p^2 3s$ levels.
The level ordering of \citet{bell00} for the $n=2$ 
configurations has been modified to be consistent with the other
nitrogen-like ions, and follows the experimental level ordering of
\citet{edlen84}. 
The new CHIANTI model for \ion{S}{10} includes lines around 50\,\AA\
due to $2s^22p^3$--$2s^22p^23s$ transitions which were not present in
earlier versions 
of the database. We have compared $n=2$ line emissivities computed
with the new  
model with the previous one \citep[based on the calculations of][]{bhat80b}
and found  differences  of the order of 20--30\%, the largest 
being for the transitions 
\orb[2s 2]\orb[2p 3] $^4S_{3/2}$ -- \orb[2s2p 4] $^4P_J$ which give
rise to emission lines between 314 and 320\,\AA.

\subsubsection{Ar\,XII, Ti\,XVI, Zn\,XXIV}

The atomic data for the $n=2$ levels of \ion{Ar}{12}, \ion{Ti}{16} and
\ion{Zn}{24} are from the same 
sources as described in  Sect.~\ref{n-zhang}: \citet{zhang99} for the
collisional data, \citet{merkelis99} and \citet{zhang99} for the
radiative data, and \citet{edlen84} for the experimental energy
levels.

Additional data for the levels in the $2s^22p^23l$ ($l=s,p,d$)
configurations have been published by \citet{bsf89}, extending the
CHIANTI models to 72 fine structure levels.
\citet{bsf89} calculated collision
strengths at a single energy for all transitions from the five levels
of the ground $2s^22p^3$ configuration. In addition they calculated
oscillator strengths and $A$ values for all allowed transitions from
the $n=3$ levels. Experimental energy levels are from the NIST
database \citep{nist}. For some levels, experimental energies were
unavailable and the theoretical energies of \citet{bsf89} were used.

\subsubsection{Fe\,XX}

The previous atomic model for \ion{Fe}{20} included the  $n=2$
\citet{bhat80a} and $n=3$ \citet{bsf89} calculations, for a total of
72 levels.  This model has been improved by including the new
$R$-matrix calculations of \citet{butler01xx}, produced as part of the
Iron Project. The authors present complete collisional and radiative
data for  86 levels within the $n=2$ and $n=3$ configurations.
Maxwellian-averaged collision strengths were tabulated for 27
temperatures between $\log\,T=5.0$ and $\log\,T=7.6$.  All of the
electron impact excitation data from the ground configuration have
been fitted with 9 point splines, making sure that  the fits were
good to $\lesssim 1$\%  over the temperature range of the original
calculation.

Experimental energy levels are from a variety of sources. For the
$2s^22p^3$ ground configuration the values of
\citet{kucera00} are used, while for all other levels in the $n=2$
configurations \citet{edlen84} is used. A number of $n=3$ level
energies have been derived from the recent line list of
\citet{brown02}. Further $n=3$ energies are from \citet{shirai00},
\citet{kelly87} and \citet{phillips99}.


The extensive line list compiled Brown et al. (2002) from laboratory
plasmas reports a number of Fe XX spectral lines that allow to determine
the energies of many levels previously unavailable. A few additional 
level energies has been identified from Phillips et al. (1999).
These new values have been used in the present version of the database 
to complement the energy values already included in the CHIANTI Version 
3 model.

The inclusion of these energies in CHIANTI allows the assignment of
laboratory-based wavelengths to many strong transitions predicted
by CHIANTI, so that these can be used for analysis.

Emissivities computed with the new \ion{Fe}{20} model show small differences
(within 20\%)  compared to the previous CHIANTI model for transitions
from the $2s2p^4$ and $2s^22p^23d$ configurations. For transitions
from the $2s^22p^23s$ configuration, however, much larger differences
of up to a factor 2 are found.

\subsection{Oxygen isoelectronic sequence}

\subsubsection{Ne\,III}

\citet{btl02} have carried out \emph{ab initio} calculations of energy
levels, radiative data and collisional excitation rates for all 
levels in the $n=2$ and \orb[2s 2]\orb[2p 2]\orb[3l ]  ($l=s,d$)
configurations, for a total of 57 fine structure levels. Collision
strengths were calculated at incident electron energies of 5, 15, 25,
35 and 45~Ry.
These calculations were
performed in the distorted wave approximation, so that the collision rates
do not include resonances. For this reason, close-coupling effective collision
strengths from \citet{mclaugh00} have been used for the 10 transitions
within the ground configuration.
The authors calculated Maxwellian-averaged collision strengths from $\log\,T=3.0$ to 6.0 at
0.2~dex intervals, and for several of the transitions it was necessary
to use 9 point splines to fit the data in order to provide coverage
over the full temperature range. \ion{Ne}{3} gives rise to several
strong forbidden lines in spectra of photoionized plasmas for which
the electron temperature is well below that in collisional ionization
equilibrium, and so it is important to provide accurate fits over a
wide temperature range.

Experimental energies are available for most of the $n=3$ levels, and
come from the NIST database \citep{nist}.

\subsubsection{S\,IX}

Several $n=3 \to n=3$ transitions have been identified in solar and
laboratory spectra in the ultraviolet range  between 700~\AA\ and
1000~\AA\ by \citet{kink97}, \citet{jupen97} and \citet{feld97}. These
lines have a great diagnostic potential if used   in conjunction with
the forbidden lines in the ground $2s^2 2p^4$ configuration found in the same
wavelength  range. Their detection has triggered the calculation of
atomic data and transition rates carried  out by \citet{bhat02b}. 
This
calculation replaces the older distorted wave calculation of
\citet{bhat79} for the $n=2$ levels used in earlier versions of CHIANTI.

\citet{bhat02b} include six different configurations in the \ion{S}{9}
atomic model: \orb[2s 2]\orb[2p 4], \orb[2s2p 5], \orb[2p 6] and
\orb[2s 2]\orb[2p 3]\orb[3l ] ($l=s,p,d$), corresponding to 86 fine
structure levels. Experimental energies are available for most of the
levels, and have been taken from the NIST database \citep{nist},
\citet{kelly87} and \citet{jupen97}; the experimental energies of two
levels have been exchanged in order  to correctly match the
theoretical levels.

Theoretical energy levels, radiative data and collision strengths are
provided by \citet{bhat02b} for all levels and transitions in the
atomic model. Collision strengths have been calculated for five values of
the incident electron energy: 25, 50, 75, 100 and 125~Ry.  Comparison
between these collision strengths and the earlier values from
\citet{bhat79}  show a reasonable agreement, although some differences
are found due to the more accurate target representation adopted by
\citet{bhat02b}.

\subsubsection{Fe\,XIX}

\citet{butler01xix} have carried out a complete calculation of energy
levels, radiative data and collisional rates for \ion{Fe}{19}, as part
of the Iron Project. This is the first calculation for \ion{Fe}{19} to
adopt the 
$R$-matrix approximation, thus enabling important resonant
contributions to be accounted for. All earlier calculations were
performed under the distorted wave approximation.

\citet{butler01xix} provide data for the \orb[2s 2]\orb[2p 4], \orb[2s2p
5], \orb[2p 6], \orb[2s 2]\orb[2p 3]\orb[3l ] ($l=s,p,d$)
configuration, as well as for the two lowest-lying triplets of the
\orb[2s2p 4]\orb[3s ] configuration. The atomic model thus includes 92
fine structure levels.

Experimental energies are taken from \citet{shirai00} and
\citet{kelly87}, but are available only for the $n=2$ and a few of the
$n=3$ levels. Recent laboratory measurements of \ion{Fe}{19} X-ray
lines \citep{brown02} allow the determination of several new level
energies, and these have been added to the CHIANTI model. An
additional energy value has also come from \citet{phillips99}.
Theoretical energies, as well as radiative data, are
provided by \citet{butler01xix} for all levels and transitions except
for the transitions within  the ground configuration, for which data
are taken from \citet{lol85}.

Maxwellian-averaged collision strengths were provided by
\citet{butler01xix} over the temperature range 
$5.0\le\log\,T\le 7.6$.
The authors claim accuracy of the $n=2$ collision rates  to
better than 20\%, but for transitions involving the $n=3$ levels
resonances coming from $n=4$ levels and beyond are omitted and so the
rates are less accurate.
Comparison between collision strengths from
\citet{butler01xix}  and the distorted wave calculations by
\citet{zhang01} and \citet{bhat89} show good agreement in the energy
regions where no resonances are present.


\subsubsection{Ni\,XXI}

Although \ion{Ni}{21} gives rise to several strong allowed
transitions, there have been no calculations of electron excitation
rates in the literature
and so it has not
been possible to include the ion in the previous versions of CHIANTI.
Recently, however, \citet{blm02}  calculated a complete set of energy
levels, radiative decay rates and collision strengths for 58 fine
structure levels from the \orb[2s 2]\orb[2p 4], 
\orb[2s2p 5], \orb[2p 6], \orb[2s 2]\orb[2p 3]\orb[3l ] ($l=s,d$)
configurations and this data-set is now included in CHIANTI.  Collision
strengths have been calculated in the 
distorted wave approximation at incident electron energies of 85, 170,
255, 340 and 425~Ry. Experimental energy levels are available for all
the $n=2$ levels and a few $n=3$  
levels, and they are taken from \citet{shirai00}.

\subsection{Fluorine isoelectronic sequence}

\subsubsection{Fe\,XVIII}

The previous \ion{Fe}{18} CHIANTI model \citep{dere97} contained
distorted wave calculations from \citet{sampson91} for all
transitions. The ground $2s^22p^5$ $^2P_{1/2}$ -- $^2P_{3/2}$ transition data
have now been replaced with the close-coupling calculations of
\citet{berr98} who tabulate Maxwellian-averaged collision strengths
for 11 temperatures over the 
range $5.5\le \log\,T \le 7.5$. The new data serve to increase the
emissivity of the \ion{Fe}{18} ground transition at \lam974.86 by
around 50\% (Fig.~\ref{fe18-fe21}). This line has recently been
observed in stellar spectra \citep{young01}.

\subsection{Magnesium isoelectronic sequence}

\subsubsection{Fe\,XV}

In the recent past, there has been a considerable effort to calculate
accurate collision rates for \ion{Fe}{15}. Independent calculations
have been made available by a number of authors, using the $R$-matrix
approach in all cases but one. \citet{eiss99}, \citet{griff99a} and
\citet{aggar99} use different variations of the same $R$-matrix
approach, while \citet{bhat97}  adopted the distorted wave
approximation. \citet{aggar00} report a comparison between the  three
$R$-matrix calculations listed above for the lowest 10 levels, and
find that the values of the collisional data for Aggarwal \etal (1999)
and \citet{eiss99} are in very good agreement, while those from
\citet{griff99a} show some unexpected problems. Corrections to the
\citet{griff99a} collisional data have been provided by the same
authors in a later note \citep{griff99b}.

In CHIANTI, we have adopted a combination of data from \citet{eiss99},
\citet{griff99a, griff99b}  and \citet{bhat97}. The atomic model includes 53
fine structure energy levels, coming from  11
configurations. Experimental energy levels come mainly from
\citet{shirai00},   although additional energies are taken from
\citet{eiss99} and \citet{read75}.

Data for the $n=3$ configurations are taken from \citet{eiss99}: these
include the 35  lowest energy levels. Theoretical energies and
effective collision strengths are provided by  the authors; however,
no radiative data is available. $A$ values and oscillator strengths
have  been calculated using SSTRUCT by one of the authors
(E. Landi) for all  53 levels in the CHIANTI model using a 21-configuration
model. Comparison of the SSTRUCT results with the \citet{griff99a}  $A$
values for $n=4$ levels yields good agreement. \citet{eiss99}  provide
Breit-Pauli $R$-matrix effective collision strengths for temperatures
in the range  $5.0\le\log\,T\le 7.0$.

Theoretical energies, radiative data and effective collision strengths
for the \orb[3s4l ] ($l=s,p,d$) levels and transitions are taken from
the corrected calculations of \citet{griff99b}. Effective collision
strengths are calculated using the ICFT $R$-matrix  method, and are
provided in the temperature range $5.05\le\log\,T\le 7.05$.

Data for the high energy \orb[3p4s ] and \orb[3s4f ] configurations
are taken from the distorted wave calculations of
\citet{bhat97}. Radiative data were computed using SSTRUCT,
while collision strengths were calculated using the University College
of London DW code for three incident electron energies: 25, 50 and
75~Ry. It is to be noted that \citet{bhat97} neglected the \orb[3d 2]
configuration and this constitutes the main deficiency of their data
for these two configurations.

\subsection{Phosphorus isoelectronic sequence}

\subsubsection{S\,II}

The collisional data of \citet{cai93} have been replaced with the more
recent results of
\citet{rams96}. These authors calculated Maxwellian-averaged collision
strengths for transitions between forty-three levels of the
$3s^2 3p^3$, $3s 3p^4$, $3s^2 3p^2 3d$, $3s^2 3p^2 4s$ and
$3s^2 3p^2 4p$ configurations of \ion{S}{2}. 
\citet{rams96} tabulate their collision strengths for temperatures
$3.5\le\log\,T\le 5.0$, however data was provided to CHIANTI by the
authors over the extended range $3.0\le\log\,T\le 6.4$. When fitting
the data over this range with the \citet{burgess92} method it was
found for a number of the allowed
transitions that the collision strengths did not tend towards the high
temperature limit point. This is due to the geometric series
approximation to the partial collision strengths failing at high
electron energies (C.A. Ramsbottom 2001, private communication). For
this reason the temperature range $3.0\le\log\,T\le 5.5$ was
considered for the fitting. Only transitions involving the five levels
of the ground $3s^23p^3$ configuration, and the metastable
$3s^23p^33d$ $^4F_{9/2}$ were fitted. 9 point splines were used  for those
data that were difficult to fit with 5 point splines,  however
it was still found necessary to omit data points from the fit in some
cases in order to obtain fits accurate to $\lesssim 1$\%. In summary
the fits in CHIANTI accurately represent all of the original data over the
temperature range $3.3\le \log\,T\le 5.1$. For many of the
transitions, however, the fits are accurate over a wider temperature range.

A complete set of radiative data for the  43 levels
of the \citet{rams96} calculations have never been
published. S. Nahar (2001, private communication) has, however,
computed radiative decay rates for all of the allowed transitions
between these levels and they have been included in the CHIANTI
model. The calculations are an extension of those presented in
\citet{nahar98}. For the forbidden transitions amongst levels in the
ground configuration and from the $3s^23p^33d$ $^4F_{9/2}$ level, the
decay rates from earlier versions of CHIANTI 
have been retained \citep{dere97}. Additional forbidden decay rates have been
computed by P.R. Young using a 24 configuration model input to the
code SSTRUCT.

Experimental energy values for all of the 43 levels in the CHIANTI
model have been obtained from the NIST database \citet{nist}.

Very significant differences are found between the present \ion{S}{2}
model and the one found in previous versions of CHIANTI. This is due
to the inaccuracies in the \citet{cai93} collisional data that have
been highlighted by \citet{rams96} and \citet{tayal97}. The new
CHIANTI model is found to provide excellent agreement with a far
ultraviolet spectrum of the Jupiter-Io torus, which hosts a large
number of $3s^23p^3$--$3s^23p^23d$ and $3s^23p^3$-$3s^23p^24s$
transitions \citep{feldman01}.

\subsubsection{Ar\,IV}

\ion{Ar}{4} is a new addition to CHIANTI, and the model includes 30
fine structure levels from the \orb[3s 2]\orb[3p 3], \orb[3s3p 4] and
\orb[3s 2]\orb[3p 2]\orb[3d ] configurations. Experimental energies
are taken from the NIST database \citep{nist} and are available for
all but two of the \ion{Ar}{4} levels.

\ion{Ar}{4} radiative data have been calculated by one of the authors
(E.~Landi) with the SSTRUCT code, using a 24-configuration atomic
model. $A$ values have been corrected for the differences between
computed and experimental energy levels; where no values were
available, original results have been retained. $A$ values and
oscillator strengths  have been compared with a number of earlier
calculations, carried out with more  limited atomic models and
different codes. Forbidden transitions within the ground
configuration have been compared with the computation of
\citet{mendoza82},  \citet{frit99} and \citet{huang84} and good
agreement was found. Optically allowed oscillator strengths have been
compared with the values from \citet{fawcett86}: the agreement is
fairly good for most transitions, although in some cases significant
differences arise.

Collisional data are taken from the $R$-matrix computations of
\citet{rams97} for transitions within the ground configuration, and
\citet{rams97a} for all other transitions. Calculations were carried
out including a 13 LS state target model, and effective collision
strengths were provided in the temperature range $3.0\le\log\,T \le
6.0$. Comparison with the earlier close-coupling calculations for
forbidden transitions in the ground  configuration from
\citet{zeipp87a} outlines significant differences for many
transitions, especially at low temperatures where resonance effects
are greatest. These  differences are probably due to the smaller
number of terms included in the \citet{zeipp87a} target
representation. Comparison of \ion{Ar}{4} lines with observations  in
optical spectra from planetary nebuale confirms the accuracy of the
adopted  collisional data \citep{keenan97}.

\subsubsection{Fe\,XII}

Recently, \citet{binello01} have reported a new set of atomic structure
calculations for \ion{Fe}{12}. The resulting theoretical energy levels and
radiative data represent an improvement relative to the earlier data
of \citet{binello98a, binello98b},
and they are adopted in the present version of CHIANTI.

\subsection{Scandium isoelectronic sequence}

\subsubsection{Fe\,VI}

The CHIANTI atomic model for \ion{Fe}{6} includes 80 fine-structure
levels, coming from the \orb[3d 3], \orb[3d 2]\orb[4s ] and \orb[3d
2]\orb[4p ] configurations. Experimental energy levels, taken from the
NIST database \citep{nist} are available for all the levels in the
adopted model.

Radiative data and theoretical energies are taken from the
SSTRUCT calculation carried out by \citet{chen00} as part of
the Iron Project. Theoretical energies and $A$ values are available
for all levels and transitions in the atomic model. Comparison of both
energies and $A$ values with results from \citet{bautista96} and
\citet{nussb78} shows good agreement.

Effective collision strengths for all possible transitions in the
adopted model have been calculated  by \citet{chen99}. The $R$-matrix
method has been used, and  effective collision strengths have been
calculated for the temperature range  $4.0\le\log\,T\le 6.0$. The
authors also investigate the importance of relativistic effects and
the effects  of numerical uncertainties associated with the resolution
of extensive resonances, finding that they are small.

\section{Continuum}

\begin{figure}[h]
\epsscale{1.0}
\plotone{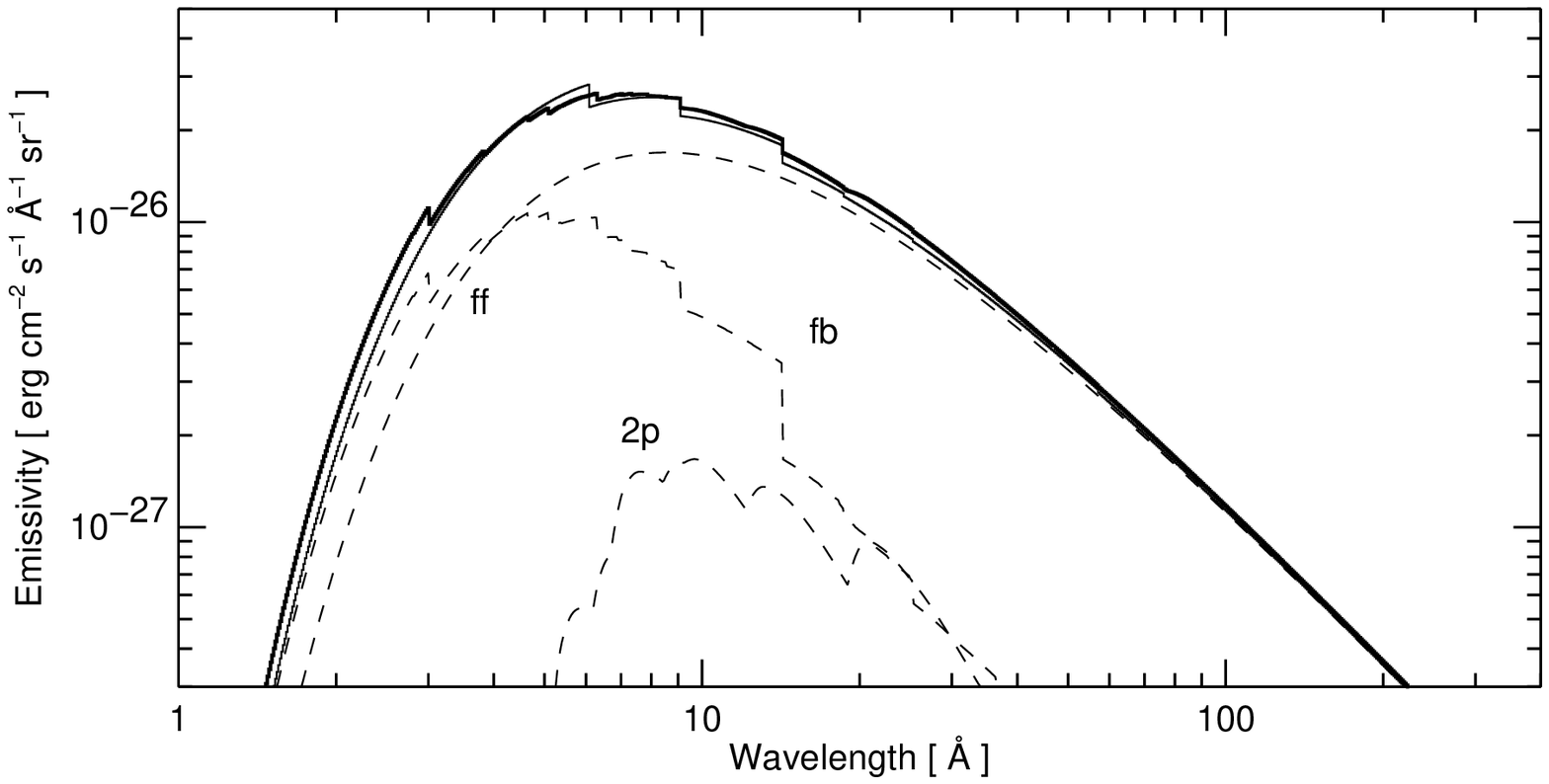}
\caption{A comparison of the continuum emissivity predicted by CHIANTI
(thick line)
at a temperature of $1\times 10^7$~K with that obtained through the
analytic approximations of \citet{mewe86}. The individual contributions of the
CHIANTI free-free (ff), free-bound (fb) and two-photon (2p) continua
are indicated 
with dashed lines.}
\label{cont-compare}
\end{figure}

An IDL routine to include the two photon continuum has been added to
CHIANTI, while the free-free (bremsstrahlung) and free-bound
(radiative recombination) continua routines have
been revised. Fig.~\ref{cont-compare} shows the total continuum
spectrum at a temperature of $1\times 10^7$~K computed for solar
photospheric abundances with the \citet{mazz98} ionization
balance. For comparison, the continuum given by the CONFLX.PRO
procedure found in the Solarsoft\footnote{Solarsoft is a set of
  integrated software libraries, data bases, and system utilities that
  provide a common programming and data analysis environment for Solar 
Physics. It is available at http://www.lmsal.com/solarsoft/.} package
is also shown. This routine 
makes use of the analytic approximations for the continua presented in
Sect.~4 of \citet{mewe86},
and is commonly used for the interpretation of solar continuum
measurements. Agreement is generally excellent with the differences
lying largely in the improved treatment of the free-bound continuum.

\subsection{Two photon continuum}

\subsubsection{Transitions in hydrogen-sequence ions}

The first excited level ($2s$ $^2S_{1/2}$) of the hydrogen
iso-electronic sequence ions can decay only by means of forbidden
magnetic dipole and two-photon transitions.  The importance of the
competing magnetic dipole transition increases with $Z$ but for nickel
($Z=28$), the two-photon transition rate is roughly 5 times that of the
magnetic dipole rate.

The spectral emissivity (erg~cm$^{-3}$ s$^{-1}$ sr$^{-1}$ \AA$^{-1}$)
for optically-thin two-photon emission at wavelength $\lambda$ is given by:
 
\begin{equation}
{d\epsilon_{i,j} \over d\lambda} =  {h c \over 4\pi \lambda}  A_{ji}
N_j(X^{+m}) \phi(\lambda_0/\lambda)
\end{equation}

\noindent
where $A_{j,i}~($sec$^{-1})$ is the Einstein spontaneous emission
coefficient ($A$ value); $N_j(X^{+m})$ is the number density of the level
$j$ of the ion $X^{+m}$; $\phi$ is the spectral distribution function;
and $\lambda_0$ is the wavelength corresponding to the energy
difference between the excited and ground level.

The transition rates for both the magnetic dipole and
two-photon transitions are taken from \citet{parpia82}.  Tables of the
spectral distribution function have been provided by \citet{goldman}
for $Z=1,20,40,60,80$ and 92.  Interpolation for values of $Z\le 28$
should be fairly accurate.

\subsubsection{Two-photon continuum transitions in helium-sequence ions}

For the helium iso-electronic sequence, the second excited level ($1s2s$
$^1S_0$) decays through a forbidden magnetic dipole and two-photon
transitions.  The two-photon decay rate has been calculated by
\citet{drake86}.  The two-photon spectral distribution has been
calculated by \citet{drakevd} for values of $Z$ between 2 and 10.  For
values of $Z$ higher than 10, we have used the spectral distribution for
$Z=10$.  The shape of the distribution function does not appear to be
changing rapidly with $Z$ at $Z=10$ so this extrapolation should be
moderately accurate.

\subsection{Bremsstrahlung}

\citet{itoh00} have provided an analytical fitting formula for the
relativistic thermal bremsstrahlung gaunt factors, and this is now
added to CHIANTI. The fitting formula is valid for the ranges
$6.0\le\log\,T\le8.5$ and $-4.0 \le\log\,(hc/k\lambda
T) \le 1.0$. For temperatures below $\log\,T=6.0$ we retain the
non-relativistic Gaunt factors of \citet{suth98} for computing the
continuum. The condition $\log\,(hc/k\lambda
T) \le 1.0$ results in some of the low wavelength points being
inaccurately represented by the Itoh et al.\ fitting formula. For
these wavelengths the Gaunt factors of \citet{suth98} are used to
compute the continuum level.

The relativistic free-free continuum is almost identical to the
non-relativistic continuum at low temperatures. At $T=1\times 10^8$~K
(the maximum temperature permitted by the ion balance calculations contained
in CHIANTI)
the relativistic continuum is around 1\% higher near the peak of the
distribution. 

The IDL procedure for calculating the bremsstrahlung emission
(FREEFREE.PRO) retains the same name and is called in an identical way
to the previous version. 

\subsection{Free-bound continuum}

The method of calculating the free-bound continuum in version~3 of
CHIANTI  was due 
to \citet{rybicki79}, setting the bound-free gaunt factors to
unity. We have revised this significantly by now 
including accurate ground photoionization cross-sections and, for
excited levels, using the gaunt factors of \citet{karzas61}.

The free-bound continuum emissivity produced from recombination onto
an ion of charge $Z$ can be written as

\begin{equation}\label{fb-eqn}
P_{{\rm fb},\lambda} =  3.0992 \times 10^{-52}
    N_{\rm e} N_{Z+1} {E_\lambda^5 \over T^{3/2}} \sum_i
    {\omega_i \over \omega_0} \sigma^{\rm bf}_i
    \exp  \left( - {E_\lambda - I_i \over kT} \right) \qquad
    [{\rm erg}~{\rm cm}^{-3}~{\rm s}^{-1}~{\rm \AA}^{-1}]
\end{equation}
where $ N_{\rm e}$ and $N_{Z+1}$ are the number densities of electrons and
recombining ions, respectively, in units of cm$^{-3}$; $E_\lambda$ is
the energy in cm$^{-1}$ of the emitted radiation; $T$ is the plasma
temperature in K; $\omega_i$ is the statistical weight of the level
$i$ in the recombined ion; $\omega_0$ is the statistical weight of the
ground level of the recombining ion; $\sigma^{\rm bf}_i$ is the
photoionization cross-section from the level $i$ in the recombined ion
to the ground level of the recombining ion, in units of Mb
($=10^{-18}$~cm$^2$); $I_i$ is the ionization energy in units of
cm$^{-1}$ from the level $i$ in the recombined ion; and $k$ is the
Boltzmann constant.
The photoionization cross-section, $\sigma^{\rm bf}_i$, is zero for
photon energies $E_\lambda < I_i$.
The sum in Eq.~\ref{fb-eqn} is over all levels $i$ below the
recombined ion's ionization limit. Within CHIANTI we take
levels  to be individual configurations within
an ion rather the usual fine structure levels employed in CHIANTI.
Accurate
photoionization cross-sections for transitions from the ground level
are readily available in the literature, and we use the analytic fits
of \citet{verner95} that are available for all ions of all elements
up to zinc.

Cross-sections for photoionizations from excited levels are generally
not available, and for these we use the hydrogenic approximation of
\citet{karzas61} where

\begin{equation}
\sigma^{\rm bf}_i = 1.075812 \times 10^{-1} {I_i^2 g_{\rm bf} \over
n_i E^3} \qquad [{\rm Mb}]
\end{equation}

\noindent where $I_i$ is the ionization energy of level $i$, $g_{\rm
bf}$ is the bound-free gaunt factor, and $n_i$ is the principal
quantum number of the ejected electron. Tables of the gaunt factors as
a function of energy for $nl$-resolved levels up to $n=6$ and $l=5$  are
published in \citet{karzas61}.

As the ion levels considered for the free-bound continuum are
different from those used in the level balance models for the ions, a
new CHIANTI file, given the suffix .FBLVL, is introduced. This
contains the configurations used for deriving the total free-bound
emissivity for the ion. E.g., for \ion{C}{3} with ground configuration
$1s^22s^22p$, we included all configurations $1s^22s^2nl$, with
$nl=2p, 3s, \ldots, 5g$. For each
configuration an energy is listed which is the weighted-average energy
of all the fine structure levels in the configuration. For the
low-lying configurations, these energies are derived from the data in
the CHIANTI .ELVLC files. For higher-lying configurations the energies
are derived in many cases from data in the NIST database. However for
some ions, particularly the iron ions, complete energy level data is
not available for $n=4,5,6$ configurations and so theoretical data
were used. Sources included atomic physics calculations already used
in CHIANTI, TOPbase\footnote{A database containing results from The
  Opacity Project \citep{seaton87}, available at \anchor{http://cdsweb.u-strasbg.fr/topbase.html}{http://cdsweb.u-strasbg.fr/topbase.html}.}, and theoretical models constructed with the
SSTRUCT atomic code.


Only the most abundant elements are considered for the calculation of
the free-bound continuum, and these are H, He, C, N, O, Ne, Mg, Al,
Si, S, Ar, Ca, Fe and Ni. The contributions from all ions in CHIANTI of these
elements are considered. Further ions not currently in CHIANTI have
also been added, including \ion{Fe}{4}, \ion{Fe}{5} and the neutrals
\ion{C}{1}, \ion{O}{1} and \ion{Si}{1}.










\section{Summary}

The previous sections have described the latest updates to the CHIANTI
atomic database that will continue to make CHIANTI a vital tool for
interpreting astrophysical data. The database and the associated IDL
software package are freely available at three websites in the US and
Europe:

\begin{itemize}
\item \anchor{http://wwwsolar.nrl.navy.mil/chianti.html}{http://wwwsolar.nrl.navy.mil/chianti.html}
\item \anchor{http://www.arcetri.astro.it/science/chianti/chianti.html}{http://www.arcetri.astro.it/science/chianti/chianti.html}
\item
  \anchor{http://www.damtp.cam.ac.uk/user/astro/chianti/chianti.html}{http://www.damtp.cam.ac.uk/user/astro/chianti/chianti.html}.
\end{itemize}

In addition, both the database and software package are available
through the Solarsoft system
(\anchor{http://www.lmsal.com/solarsoft/}{http://www.lmsal.com/solarsoft/}). 

\acknowledgements

We are grateful to our colleagues who have provided their data to
us. These include A. K. Bhatia, S. Nahar,  S. Tayal, and H. L. Zhang.
We thank R.K.~Smith for providing many proton rate data-sets in
electronic format.  We also acknowledge helpful discussions with
C. Jordan and M. Copetti.  The work of K. P. Dere and E. Landi was
supported by NASA's Applied Information Systems Research Program and
by a grant from the Chandra Emission Line Project. G.\ Del Zanna and
H. E.\ Mason acknowledge  support from PPARC.  The CHIANTI team has
benefited from a travel grant from NATO.

\end{document}